\documentclass[a4paper,11pt]{article}
\usepackage{geometry}
\usepackage{a4wide,slashbox}
\usepackage{graphicx}
\usepackage{epsf}
\usepackage{amsmath}
\usepackage{amssymb}

\usepackage{cite}
\usepackage{multirow,tabularx}
\usepackage{appendix}
\usepackage{graphicx,amsmath,amsfonts,amssymb,amsthm,euscript,braket,xcolor}
\newcommand{\be}{\begin{equation}}
\newcommand{\ee}{\end{equation}}

\newcommand{\Rmnum}[1]{\expandafter\@slowromancap\romannumeral #1@}
\newcommand{\bea}{\begin{eqnarray}}
\newcommand{\eea}{\end{eqnarray}}

\numberwithin{equation}{section}

\begin{document}

\title{\bf Thermal entropy of a quark-antiquark pair above and below deconfinement from a dynamical holographic QCD model}
\author{\textbf{David Dudal$^{a,b}$}\thanks{david.dudal@kuleuven.be},
\textbf{Subhash Mahapatra$^{a}$}\thanks{subhash.mahapatra@kuleuven.be}
 \\\\\
\textit{{\small $^a$ KU Leuven Kulak, Department of Physics, Etienne Sabbelaan 53 bus 7657,}}\\
\textit{{\small 8500 Kortrijk, Belgium}}\\
\textit{{\small $^b$           Ghent University, Department of Physics and Astronomy, Krijgslaan 281-S9, 9000 Gent, Belgium}}}
\date{}

\maketitle
\abstract{We discuss the entropy carried by a quark-antiquark pair, in particular across the deconfinement transition. We therefore rely on a self-consistent solution to Einstein-Maxwell-dilaton gravity, capable of mimicking essential features of QCD. In particular we introduce a novel model that still captures well the QCD confinement and deconfinement phases, while allowing the introduction of a temperature in a phase which resembles the confined phase, this thanks to it being dual to a small black hole. We pay due attention to some subtleties of such model. We confirm the lattice picture of a strong build-up of thermal entropy towards the critical temperature $T_c$, both coming from below or above $T_c$. We also include a chemical potential, confirming this entropic picture and we consider its effect on the speed of sound. Moreover, the temperature dependent confinement phase from the holography side allows us to find a string tension that does not vanish at $T_c$, a finding also supported by lattice QCD. }

\section{Introduction}
The investigation of QCD at finite temperature is extremely important for various reasons, ranging from heavy ion physics to cosmology. It is expected that studies of heavy quarkonium at finite temperature may enhance our understanding of the deconfinement transition and that these may shed new light onto the (creation of the) QCD plasma phase. A lot of new results, both from experimental as well as from theoretical side, are appearing. For example, recent experimental results from RHIC and LHC have indicated a strong suppression of charmonium near the deconfinement transition \cite{Adare,Abelev:2013ila}. The non-trivial experimental result that the charmonium suppression is stronger at lower energy density than at larger energy density has led to an intense investigation into the nature of this suppression and many theoretical scenarios such as the renown colour screening mechanism \cite{Matsui:1986dk}, a recombination of the produced charm quarks into charmonia \cite{BraunMunzinger:2000px,Thews:2000rj}, the imaginary potential mechanism for heavy quark dissociation \cite{Laine:2006ns,Escobedo:2014gwa,Noronha:2009da}, etc have been suggested for its explanation.\\

\begin{figure}[t!]
\begin{minipage}[b]{0.5\linewidth}
\centering
\includegraphics[width=2.8in,height=2.3in]{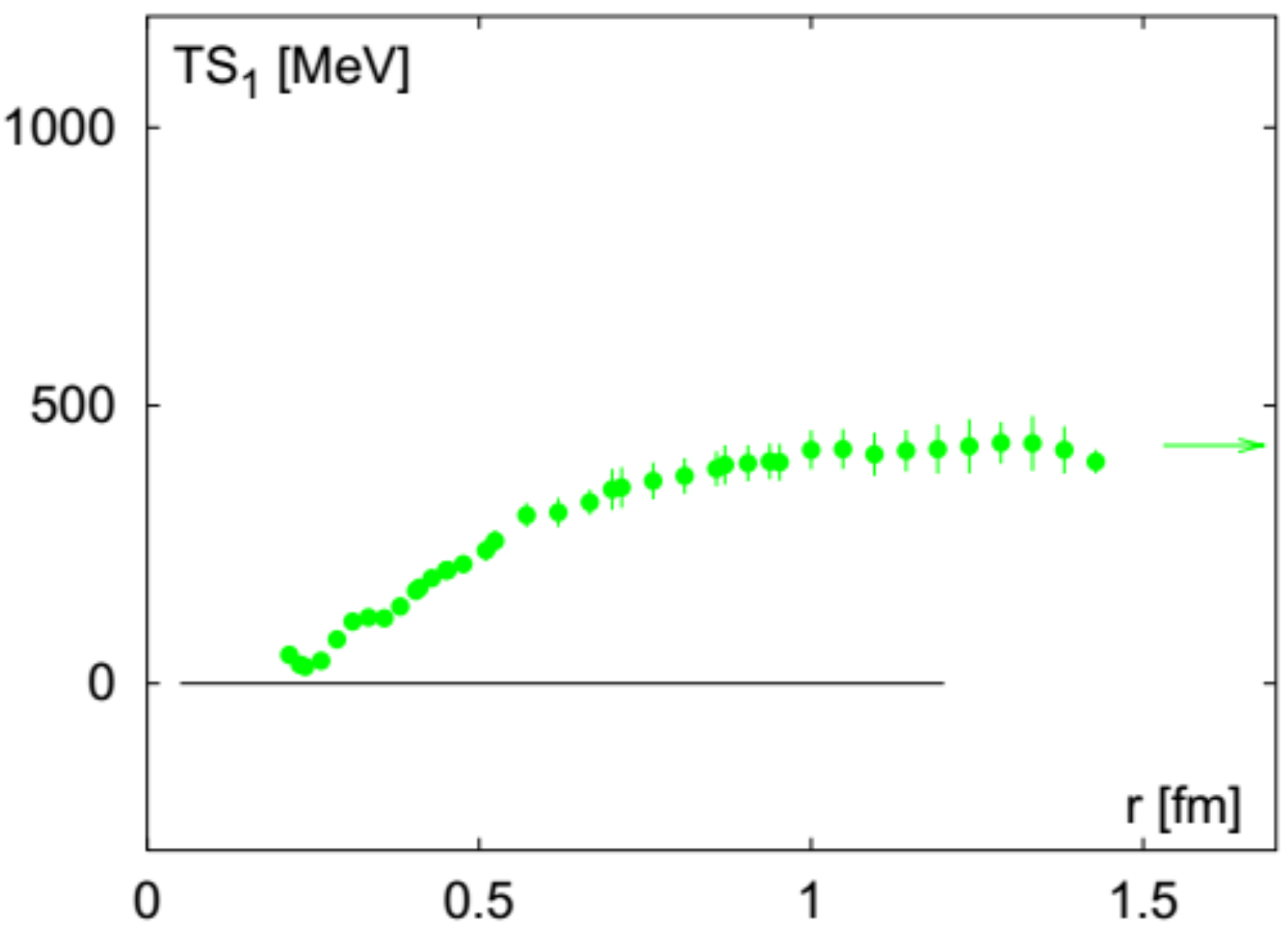}
\caption{ \small Two flavour lattice QCD result for the entropy of a $q\overline q$ pair as function of quark-antiquark separation at temperature $T\simeq 1.3T_c$. The figure is taken from \cite{Kaczmarek:2005zp}.}
\label{latticeQCD2}
\end{minipage}
\hspace{0.4cm}
\begin{minipage}[b]{0.5\linewidth}
\centering
\includegraphics[width=2.8in,height=2.5in]{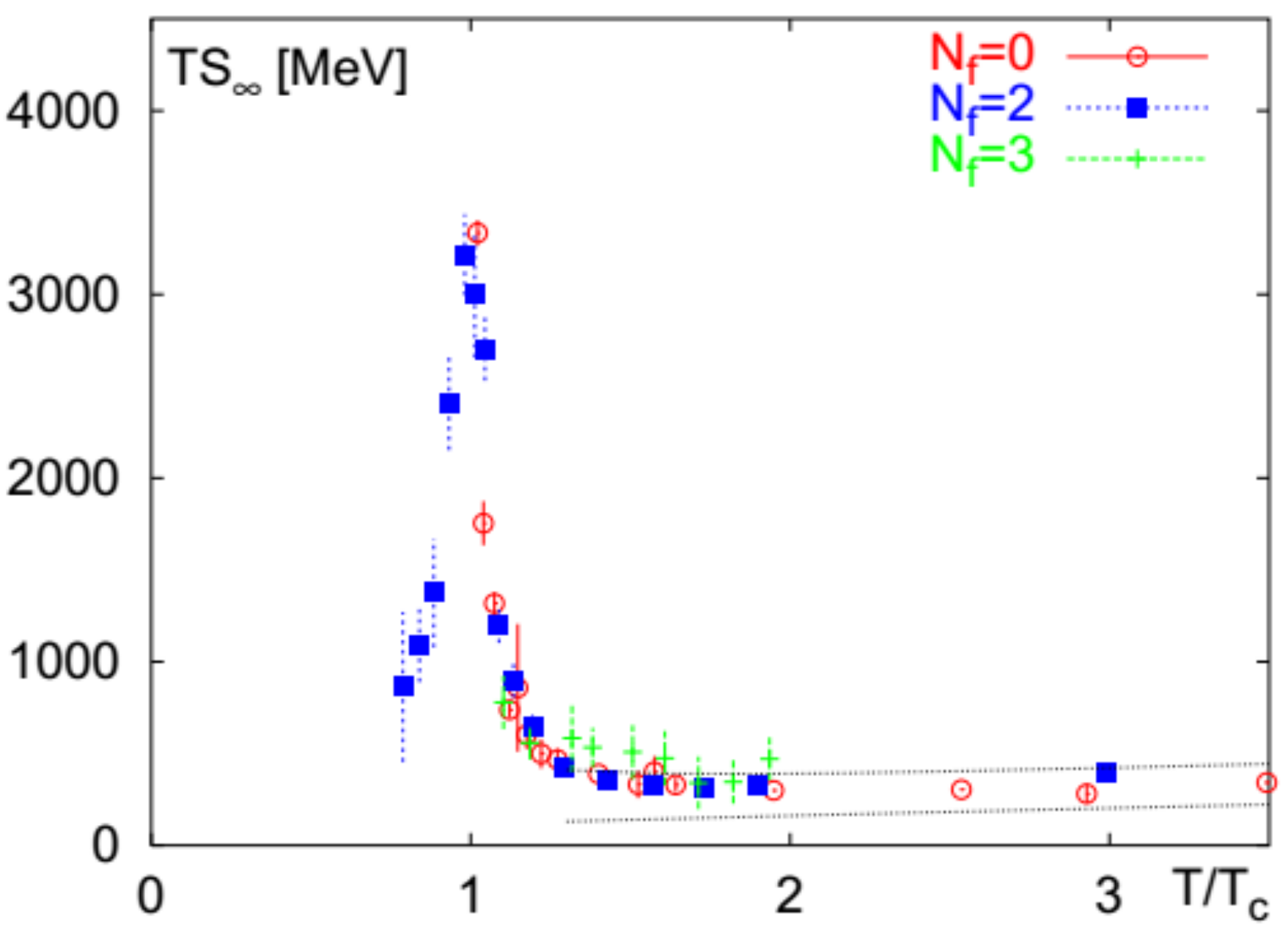}
\caption{\small Lattice QCD result for the entropy of a $q\overline q$ pair as function of temperature $T/T_c$ for large $q\overline q$ separation. The figure is taken from \cite{Kaczmarek:2005zp}. }
\label{latticeQCD1}
\end{minipage}
\end{figure}

One such mechanism which attracted interest of late, and a topic of this paper, was proposed in \cite{Kharzeev:2014pha}: it was suggested that the suppressed production of heavy quarkonium may be related to the nature of the deconfinement transition. In \cite{Kharzeev:2014pha}, based on lattice QCD results \cite{Kaczmarek:2002mc,Petreczky:2004pz,Kaczmarek:2005zp}, the notion of an ``emergent entropic force'', $$F=T\frac{\partial S}{\partial \ell},$$ was introduced. Here $S$ denotes the entropy, $T$ is the temperature and $\ell$ is the inter-quark separation. The lattice prediction that the entropy of a quark-antiquark ($q\overline q$) pair increases with distance, shown in Figure~\ref{latticeQCD2}, leads to a positive repulsive force and can promote the self-destruction of the bound state. It was shown that the destruction of a bound state to a delocalized state was maximal near the deconfinement transition temperature, giving a possible explanation for the strong suppression of quarkonium around that temperature.\\

The emergent entropic force mechanism of \cite{Kharzeev:2014pha} is thus based on lattice QCD simulations which suggested a considerable amount of entropy is associated with a $q\overline q$ pair in the hot QCD plasma, indicating a strong entanglement between the heavy bound states with the rest of QCD plasma. The lattice QCD data are summarized in Figures~\ref{latticeQCD2} and \ref{latticeQCD1}. Some of the main lattice observations are i) prediction of a large amount of entropy associated with a heavy $q\overline q$ pair near the deconfinement transition temperature ii) away from the deconfinement temperature the entropy decreases with temperature in both confined and deconfined phases and iii) the entropy grows as the separation between the quarks increases. This entropy first increases gradually and then saturates to a constant value at large separations.\\
\\
As emphasized in \cite{Kaczmarek:2005ui,Hashimoto:2014fha}, the lattice data for the $q\overline q$ entropy at the peak near the transition temperature also indicates the breakdown of the weak coupling approximation. The entropy near the transition temperature computed from a mean field Debye screening approximation was found to be an order of magnitude lower than that of lattice results. This suggests the need of a method which is more suitable for entropy calculations in the strongly coupled regime. One such non-perturbative method is the gauge/gravity duality \cite{Maldacena:1997re,Gubser:1998bc,Witten:1998qj}. Using this duality, a lot of new insights into the regime of strongly coupled QCD, which agrees qualitatively with the lattice QCD results, have been obtained both from ``top-down'' and phenomenological ``bottom-up'' models  \cite{Witten9803,Polchinski0003,Sakai0412,Sakai0507,Kruczenski0311,Karch0205,Klebanov0007,Erlich,GursoyI,GursoyII,Gursoy,Gursoy:2010fj,Gursoy:2009jd,Alho:2012mh,Alho:2013hsa,Jarvinen:2015ofa,Herzog0608,Karch0602,
Karch1012,Colangelo:2011sr,Dudal:2015wfn,Dudal:2014jfa,Dudal:2015kza,Callebaut:2011ab,Gherghetta:2009ac,Panero:2009tv}. Since gauge/gravity duality provides a valuable and unique technique to investigate strongly coupled gauge theory, therefore it would be interesting to see whether this duality can also provide new insights into the entropy of the $q\overline q$ pair.\\

The entropy of the $q\overline q$ pair and the idea of an entropic force was first discussed holographically in \cite{Hashimoto:2014fha} for top-down models. Two models --- $\mathcal{N}=4$ supersymmetric Yang-Mills theory (SYM) and Pure 4D Yang-Mills theory obtained from compactification of 5D SYM on a circle--- were considered and the holographic $q\overline q$ thermal entropy was calculated from the free energy of an open string hanging from the asymptotic boundary into the bulk spacetime \cite{Maldacena:1998im,Rey:1998ik,Rey:1998bq}. Albeit not exactly corresponding to the (lattice) QCD setup, similar features were seen in these dual black hole phases: the growth of the entropy with inter-quark distance, saturation to a constant value at large separations and in the second case, a divergence at the deconfinement temperature, reflecting the sharp peak observed in genuine lattice QCD. These calculations were then extended to other holographic models in \cite{Iatrakis:2015sua,Fadafan:2015ynz,Zhang:2016fwr}. However, since these two holographic models do not accurately describe real QCD, there were some discrepancies too. For example, the higher temperature asymptotics of the entropy do not reconcile with lattice QCD results. This discrepancy was later rectified in \cite{Iatrakis:2015sua} using the improved holographic bottom-up models of \cite{GursoyI,GursoyII}. Another important lattice result which could not be reproduced by holographic models in \cite{Hashimoto:2014fha,Iatrakis:2015sua} was the temperature dependence of the entropy in the confined phase. This is precisely due to the fact that in the gauge/gravity duality, the confined phase is generally dual to an asymptotically AdS space (without a horizon), which is thence independent of temperature. Subsequently all the physical quantities in the confined phase are independent of temperature. In order to study temperature dependence in this phase one would need to consider $1/N$ corrections in the holographic models, something extremely difficult to calculate.\\

However there might be another way to introduce temperature dependence in the confined phase, without having to compute $1/N$ corrections: we can construct a black hole solution whose dual boundary theory satisfies all the necessary properties of confinement. By confinement we simply mean a phase that satisfies area law behaviour for the (expectation value of the) Wilson loop and a Polyakov loop with vanishing expectation value~\footnote{We work in the holographic analogue of quenched QCD, i.e.~with non-propagating quark flavour degrees of freedom. As such, QCD enjoys an explicit $\mathbb{Z}_N$ symmetry in the confined phase, with the Polyakov loop expectation value serving as order parameter.}. Indeed, one can see from Figure~\ref{latticeQCD1} that even in the confined phase a large amount of entropy is associated with the $q\overline q$ pair and therefore it is important to try to analyze this observation from the holography viewpoint. The black hole solution will naturally introduce a notion of temperature in the confined phase and it would be interesting to see whether such solution, if constructed, can also capture lattice results for the entropy of the $q\overline q$ pair. We will show in section 3 that it is indeed possible to construct such a close cousin of the confined phase, which we call \textit{specious-confined} phase, for which the dual gravity side contains a black hole. Interestingly, the entropy of the $q\overline q$ pair in this holographic \textit{specious-confined} phase turns out to be in qualitative agreement with lattice QCD confined phase results. The \textit{specious-confined} model will also allow us to study, to our knowledge for the first time from holography, the temperature dependence of the ensuing QCD string tension, finding qualitative agreement with quenched lattice QCD as well, in the sense that the string tension does not vanish at deconfinement \cite{Kaczmarek:1999mm,Cardoso:2011hh,Cea:2015wjd}. \\

It is also well known that the QCD phase diagram strongly depends on the chemical potential (i.e.~a nonzero quark density). In particular, the confinement/deconfinement transition temperature and QCD equation of state are sensitive to the value of the chemical potential. Accordingly, one might think that the entropy of the $q\overline q$ pair may also depend non-trivially on it. However such results are are not available in lattice QCD yet due to the sign problem plaguing finite density simulations. In the absence of lattice QCD results, therefore, it is of great importance to investigate the entropy holographically and see the effects of chemical potential on it.\\

In this work our aim is to construct one such holographic QCD model that displays qualitative agreement with lattice results for the $q\overline q$ entropy, not only for the deconfined phase but in particular for the confined phase as well. We will also include effects of a chemical potential. For this purpose we consider a phenomenological bottom-up Einstein-Maxwell-dilaton gravity model  \cite{Gubser:2008yx,Gubser:2008ny,DeWolfe:2011ts,Rougemont:2015ona,Cai:2012xh,Noronha:2009ud,Noronha:2010hb,He:2013qq,Yang:2015aia,
Finazzo:2013aoa,Knaute:2017opk}.  The gravity system is analytically solvable in terms of an arbitrary scale function $A(z)$ (see eq.~(\ref{metsolution1})), depending on which there can be various kinds of phase transitions on the gravity side. In this paper, we consider two different forms for $A(z)$. The first form, discussed in eqs.~(\ref{Aansatz1}), leads to a first order phase transition in terms of a small to a large black hole phase. These small and large black hole phases correspond to \textit{specious-confinement} and deconfinement phases in the dual boundary theory respectively. Notably, the boundary dual of the small black hole phase does not exactly correspond to confinement as it has a non-zero (albeit exponentially small) Polyakov loop expectation value while showing linear confinement for larger distances at low temperatures only. For this reason we called this dual phase \textit{specious-confined} instead of confined. Furthermore, the \textit{specious-confinement}/deconfinement transition temperature decreases with the chemical potential and the corresponding first order small/large black hole phase transition line terminates at a second order critical point, as predicted by the well-known lattice study \cite{deForcrand:2002hgr}. Similarly, the second form of $A(z)$, shown in eqs.~(\ref{Aansatz2}), leads to a first order Hawking/Page phase transition from thermal-AdS to black hole on the gravity side. These thermal-AdS/black hole phases correspond to the standard confinement/deconfinement phases in the dual boundary QCD theory.\\

Having constructed a black hole solution (using the first form of $A(z)$) for both \textit{specious-confinement} and deconfinement phases, we then study temperature dependence of the $q\overline q$ entropy in these phases. We find that our phenomenological model does qualitatively reproduce the lattice QCD results of the $q\overline q$ entropy. Our main results are shown in Figures~\ref{TvsSvsMuconfdcase1}, \ref{TvsSvsMudeconfdcase1} and \ref{lvsSvsTMu0case1}. In the deconfined phase, we find that our gravity model displays the growth of the entropy with inter-quark distance, corroborating the entropic force scenario of \cite{Kharzeev:2014pha} for the self-destruction of the bound state. This entropy saturates to a constant value at large distances, as also predicted by lattice QCD. Our model further reproduces a sharp peak in the entropy near the critical temperature. The sharp peak near the critical temperature can be appreciated from both deconfined as well as from \textit{specious-confined} phases. Interestingly, even though \textit{specious-confined} is strictly speaking not referring to a genuine confined phase, the entropy of the $q\overline q$ pair matches qualitatively with the lattice QCD predictions.  We further provide a holographic estimate for the $q\overline q$ entropy with chemical potential which could be tested using lattice QCD in the near future. Similarly, the results for the $q\overline q$ entropy using the second form of $A(z)$ are again in agreement with lattice QCD. The qualitative results in the deconfined phase remains the same, however, since the confined phase is now dual to thermal-AdS, the entropy is zero in that phase.\\

We also study the speed of sound ($C_{s}^2$) in our holographic models and find results which again are in qualitative agreement with lattice QCD. In particular, as in lattice QCD, $C_{s}^2$ is greatly suppressed near the transition temperature in both models. Importantly, $C_{s}^2$ is suppressed in the \textit{specious-confined} phase side too. We find that $C_{s}^2$ approaches its conformal value $1/3$ from below at high temperatures. Our results further indicate that higher values of chemical potential try to suppress $C_{s}^2$ more near the transition temperature. However for extremely low (in the \textit{specious-confined} phase) and high temperatures, $C_{s}^2$ approaches a chemical potential independent constant value. \\

The paper is organized as follows. In the next section, we describe our gravity model and obtain the solution analytically. In section 3, using a particular form of $A(z)$, we first study the thermodynamics of the gravity solution and then discuss the free energy and entropy of a $q\overline q$ pair with  and compare it with lattice QCD results. We repeat the calculations of section 3 with a different $A(z)$ Ansatz in section 4. In section 5, we study the speed of sound in the constructed \textit{specious-confined}, confined and deconfined phases. Finally, we conclude this paper with some discussions and an outlook to future research in section 6.

\section{Einstein-Maxwell-dilaton gravity}
We start with the Einstein-Maxwell-dilaton action in five dimensions,
\begin{eqnarray}
&&S_{EM} =  -\frac{1}{16 \pi G_5} \int \mathrm{d^5}x \sqrt{-g} \ \ \bigl[R-\frac{f(\phi)}{4}F_{MN}F^{MN} -\frac{1}{2}\partial_{M}\phi \partial^{M}\phi -V(\phi)\bigr]
\label{actionEF}
\end{eqnarray}
where $G_5$ is the corresponding Newton constant, $V(\phi)$ is the potential of the dilaton field and $f(\phi)$ is a gauge kinetic function which represents the coupling between dilaton and gauge field $A_{M}$. The Einstein, Maxwell and dilaton equations of motion derived from eq.~(\ref{actionEF}) are, respectively,
\begin{eqnarray}
R_{MN}-\frac{1}{2}g_{MN}R-T_{MN}=0,
\label{EinsteinEOM}
\end{eqnarray}
\begin{eqnarray}
\nabla_{M}[f(\phi) F^{MN}]=0,
\label{MaxwellEOM}
\end{eqnarray}
\begin{eqnarray}
\partial_{M} \bigl[ \sqrt{-g}\partial^{M}\phi \bigr]-\sqrt{-g} \biggl( \frac{\partial V}{\partial \phi} + \frac{F^2}{4}\frac{\partial f}{\partial \phi} \biggr)=0
\label{dilatonEOM}
\end{eqnarray}
where
$$T_{MN}=\frac{1}{2} \biggl(\partial_{M}\phi \partial_{M}\phi-\frac{1}{2}g_{MN} (\partial \phi)^2 -g_{MN}V(\phi) \biggr)+\frac{f(\phi)}{2}\biggl(F_{MP}F_{N}^{\ P} -\frac{1}{4}g_{MN}F^2 \biggr).$$
In order to simultaneously solve eqs.~(\ref{EinsteinEOM}), (\ref{MaxwellEOM}) and (\ref{dilatonEOM}), we consider the following Ans\"atze for metric, gauge and dilaton fields,
\begin{eqnarray}
& & ds^2=\frac{L^2 e^{2 A(z)}}{z^2}\biggl(-g(z)dt^2 + \frac{dz^2}{g(z)} + dy_{1}^2+dy_{3}^2+dy_{3}^2 \biggr)\,, \nonumber \\
& & A_{M}=A_{t}(z), \ \ \ \ \phi=\phi(z)
\label{metric}
\end{eqnarray}
where we have assumed that the various fields depend only on the extra radial coordinate $z$. Here $L$ is the AdS length scale and in our notation, $z=0$ corresponds to the asymptotic boundary of the spacetime, i.e.~where the strongly coupled gauge theories is located.\\

Plugging the above Ans\"atze into eqs.~(\ref{EinsteinEOM})-(\ref{dilatonEOM}), we get the following equations
\begin{eqnarray}
\phi'' +\phi' \left( -\frac{3}{z}+\frac{g'}{g}+3A' \right)- \frac{L^2 e^{2A}}{z^2 g} \frac{\partial V}{\partial \phi}+\frac{z^2 e^{-2A}A_{t}'^{2}}{2 L^2 g} \frac{\partial f}{\partial \phi}=0,
\label{phiEOM}
\end{eqnarray}
\begin{eqnarray}
A_{t}'' +A_{t}' \left( -\frac{1}{z}+\frac{f'}{f}+A' \right)=0,
\label{AtEOM}
\end{eqnarray}
\begin{eqnarray}
g'' + g'\biggl ( -\frac{3}{z} + 3A' \biggr)- \frac{e^{-2A} A_{t}'^{2} z^2 f}{L^2}=0,
\label{fEOM}
\end{eqnarray}
\begin{eqnarray}
A''+\frac{g''}{6g} + A' \left(-\frac{6}{z}+\frac{3g'}{2g}\right)-\frac{1}{z}\left(-\frac{4}{z}+\frac{3g'}{2g}\right)
+3 A'^{2} + \frac{L^2 e^{2A} V}{3z^2 g}=0,
\label{AEOM1}
\end{eqnarray}
\begin{eqnarray}
A''- A'\left(-\frac{2}{z}+A'\right)+\frac{\phi'^2}{6}=0
\label{AEOM2}
\end{eqnarray}
where only four of the above five equations are independent. In the following, we treat eq.~(\ref{phiEOM}) as a constraint equation and consider eqs.~(\ref{AtEOM})-(\ref{AEOM2}) as independent equations. Importantly, these independent equations can be solved analytically in term of functions $A(z)$ and $f(z)$ i.e.~in terms of a scaling factor and a kinetic gauge function \cite{He:2013qq,Yang:2015aia}. In term of $A(z)$ and $f(z)$, the solutions are
\begin{eqnarray}
&&g(z)=1-\frac{\int_{0}^{z} dx \  x^3 e^{-3A(x)} \int_{x_c}^{x} dx_1 \  \frac{x_1 e^{-A(x_1)}}{f(x_1)}}{\int_{0}^{z_h} dx \ x^3 e^{-3A(x)}  \int_{x_c}^{x} dx_1 \  \frac{x_1 e^{-A(x_1)}}{f(x_1)} }, \nonumber \\
&&\phi'(z)=\sqrt{6(A'^2-A''-2 A'/z)}, \nonumber \\
&& A_{t}(z)=\sqrt{\frac{-1}{\int_{0}^{z_h} dx \ x^3 e^{-3A(x)}  \int_{x_c}^{x} dx_1 \  \frac{x_1 e^{-A(x_1)}}{f(x_1)}}}   \int_{z_h}^{z} dx \  \frac{x e^{-A(x)}}{f(x)}, \nonumber \\
&&V(z)=-\frac{3z^2ge^{-2A}}{L^2}\bigl[A''+A' \bigl(3A'-\frac{6}{z}+\frac{3g'}{2g}\bigr)-\frac{1}{z}\bigl(-\frac{4}{z}+\frac{3g'}{2g}\bigr)+\frac{g''}{6g} \bigr]
\label{metsolution}
\end{eqnarray}
where we have used the boundary condition that at the horizon $g(z_h)=0$ and $g(z)$ goes to 1 at the asymptotic boundary. It can be explicitly verified, with the help of the potential expression in eq.~(\ref{metsolution}), that the Einstein, Maxwell and dilaton equations are indeed satisfied. The undetermined integration constant $x_c$ in eq.~(\ref{metsolution}) can be fixed in terms of a chemical potential ($\mu$) present in the boundary theory. Expanding $A_{t}$ near the asymptotic boundary $z=0$ and using the gauge/gravity mapping, we get
\begin{eqnarray}
\mu=-\sqrt{\frac{-1}{\int_{0}^{z_h} dx \ x^3 e^{-3A(x)}  \int_{x_c}^{x} dx_1 \  \frac{x_1 e^{-A(x_1)}}{f(x_1)}}}   \int_{0}^{z_h} dx \  \frac{x e^{-A(x)}}{f(x)}.
\label{Mueqn}
\end{eqnarray}
The only non-trivial inputs which remain to be fixed are $A(z)$ and $f(z)$. Indeed eq.~(\ref{metsolution}) is a gravity solution for the Einstein-Maxwell-dilaton system for any $A(z)$ and $f(z)$. Therefore, we have a freedom of choosing any $A(z)$ and $f(z)$. We can use this freedom to constrain the arbitrary factors $A(z)$ and $f(z)$ by matching the properties of the boundary gauge theory with real QCD. For example, $f(z)$ can be fixed by demanding the dual boundary theory to satisfy linear Regge trajectories, i.e.~the squared mass of the mesons vary linearly with respect to the radial excitation number. In this paper, we will consider the following simple form for $f(z)$
\begin{eqnarray}
f(z)=e^{-c z^2 -A(z)}.
\label{fansatz}
\end{eqnarray}
Using this form of $f(z)$, one can easily show that the discrete spectrum of the mesons indeed lie on a linear Regge trajectory. One can fix the magnitude of $c$ by matching the holographic meson mass spectrum to that of lowest lying heavy meson states \cite{He:2013qq}. By doing that, one gets~\footnote{The factor $c$ can also be fixed by matching with the $\rho$ meson mass instead, as done in e.g.~\cite{Andreev:2006vy}. However, since we are more interested in the properties of heavy bound states, we find it more appropriate to compare with the $J/\Psi$ and $\Psi'$ state, following \cite{He:2013qq}.} $$c=1.16 \ \text{GeV}^2.$$
With this choice of $f(z)$, the gravity solution is then given by
\begin{eqnarray}
&&g(z)=1-\frac{1}{\int_{0}^{z_h} dx \ x^3 e^{-3A(x)}} \biggl[\int_{0}^{z} dx \ x^3 e^{-3A(x)} + \frac{2 c \mu^2}{(1-e^{-c z_{h}^2})^2} \det \mathcal{G}  \biggr],\nonumber \\
&&\phi'(z)=\sqrt{6(A'^2-A''-2 A'/z)}, \nonumber \\
&& A_{t}(z)=\mu \frac{e^{-c z^2}-e^{-c z_{h}^2}}{1-e^{-c z_{h}^2}}, \nonumber \\
&&V(z)=-\frac{3z^2ge^{-2A}}{L^2}\left[A''+A' \left(3A'-\frac{6}{z}+\frac{3g'}{2g}\right)-\frac{1}{z}\left(-\frac{4}{z}+\frac{3g'}{2g}\right)+\frac{g''}{6g} \right]
\label{metsolution1}
\end{eqnarray}
where
\[
\det \mathcal{G} =
\begin{vmatrix}
\int_{0}^{z_h} dx \ x^3 e^{-3A(x)} & \int_{0}^{z_h} dx \ x^3 e^{-3A(x)- c x^2} \\
\int_{z_h}^{z} dx \ x^3 e^{-3A(x)} & \int_{z_h}^{z} dx \ x^3 e^{-3A(x)- c x^2}
\end{vmatrix}.
\]

We now have a complete solution of Einstein-Maxwell-dilaton gravity system. The solution in eq.~(\ref{metsolution1}) corresponds to a black hole with horizon at $z_h$. The Hawking temperature and entropy of this black hole solution are given by,
\begin{eqnarray}
T&=& \frac{z_{h}^3 e^{-3 A(z_h)}}{4 \pi \int_{0}^{z_h} dx \ x^3 e^{-3A(x)}} \biggl[ 1\nonumber\\&&+\frac{2 c \mu^2 \bigl(e^{-c z_h^{2}}\int_{0}^{z_h} dx \ x^3 e^{-3A(x)}-\int_{0}^{z_h} dx \ x^3 e^{-3A(x)}e^{-c x^{2}} \bigr)}{(1-e^{-c z_h^{2}})^2} \biggr] \nonumber \\
S_{BH}&=& \frac{L^3 e^{3 A(z_h)}}{4 G_5 z_{h}^3}.
\label{Htemp}
\end{eqnarray}
Let us also mention that there is another solution to the Einstein-Maxwell-dilaton equations, i.e.~one without a horizon corresponding to a thermal-AdS space. The thermal-AdS solution can be obtained by taking the $z_h\rightarrow\infty$ limit of the black hole solution presented above, i.e.~$g(z)=1$. Although this thermal solution again asymptotes to AdS at the boundary, however depending on the scale factor $A(z)$, it can have a non-trivial structure in the bulk. This solution will play a significant role in our analysis in section 4.\\

The only function that now remains to be fixed is the scale factor $A(z)$. Indeed, in order to proceed further we now need to specify a specific form for $A(z)$. In the next two sections, we consider two different forms of $A(z)$ and study the thermodynamics of the gravity system, the free energy and entropy of the $q\overline q$ pair, speed of sound \textit{etc} for each form separately. As we will show, depending on the form of $A(z)$, there can be various kinds of phase transitions between the gravity solutions, and correspondingly, the interpretation of these gravity solutions in the boundary gauge theory can be different.\\

Above we have presented a gravity solution in the Einstein frame, useful for investigating thermodynamical properties and equations of state of the system. However, in order to study some confinement/deconfinement properties and in particular the free energy of a $q\overline q$ pair, it is more convenient to use the string frame metric. It can be obtained from the Einstein frame metric by the following standard transformation via the dilaton,
\begin{eqnarray}
& & (g_s)_{MN}=e^{\sqrt{\frac{2}{3}}\phi} g_{MN}, \nonumber \\
& & ds_{s}^2= \frac{L^2 e^{2 A_{s}(z)}}{z^2}\biggl(-g(z)dt^2 + \frac{dz^2}{g(z)} + dy_{1}^2+dy_{3}^2+dy_{3}^2 \biggr)
\label{stringmetric}
\end{eqnarray}
where $A_{s}(z)=A(z)+\sqrt{\frac{1}{6}} \phi(z)$.\\

For completeness, we should mention here that we treat the scalar field $\phi$ as a genuine dilaton, although our models are lacking, as any bottom-up AdS/QCD model, a proper embedding into an underlying string theory. If $\phi$ was just another scalar field other than a dilaton~\footnote{See for example \cite{Kiritsis:2009hu,DeWolfe:2009vs,Charmousis:2010zz} exploring this possibility.} so that $(g_s)_{MN}\equiv g_{MN}$, then confinement and all of its related features are absent and thus any qualitative comparing with QCD becomes obsolete, both from the Case 1 and Case 2 model we are going to discuss below. This finding is analogous to another self-consistent model used to connect to magnetized QCD, as introduced and discussed in \cite{Critelli:2016cvq}.\\

Before ending this section, we like to mention that the Gubser criterion of \cite{Gubser:2000nd} for Einstein-scalar gravity theory is satisfied in our gravity model as well~\footnote{Other, stronger, criteria for the physical reliability of generic naked singularities are suggested in \cite{Charmousis:2010zz,Gouteraux:2012yr}, but our potentials fall beyond the class discussed there.}. More precisely, the dilaton potential at any point in the bulk is bounded from above by its value at the UV boundary, i.e.~$ V(0)\geq V(z)$.

\section{Case 1: specious confinement}
Let us first consider the following simple form of $A(z)$,
\begin{eqnarray}
A(z)=A_{1}(z)=-\frac{3}{4}\ln{(a z^2+1)}+\frac{1}{2}\ln{(b z^3+1)}-\frac{3}{4}\ln{(a z^4+1)}.
\label{Aansatz1}
\end{eqnarray}
It is straightforward to see that $A_{1}(z)\rightarrow 0$ at the boundary $z=0$. This is nothing but the statement that bulk spacetime asymptotes to AdS at the boundary. Moreover, expanding the dilaton field and potential near the asymptotic boundary and rewriting the dilaton potential in terms of dilaton field, we get
\begin{eqnarray}
V(\phi)=-\frac{12}{L^2}+\frac{\Delta(\Delta-4)}{2}\phi^2(z)+\ldots, \ \ \Delta=3.
\label{Vcase1exp}
\end{eqnarray}
The above equation indicates that the dilaton mass $m^2$ meets a well known result in gauge/gravity duality, i.e.~$m^2=\Delta(\Delta-4)$. The dimension $\Delta=3$ of the dual operator satisfies the Breitenlohner-Freedman (BF) bound $2<\Delta<4$ as well \cite{Breitenlohner:1982bm,Breitenlohner:1982jf}. Furthermore, near the boundary we have $V(z)|_{z\rightarrow 0}=-12/L^2=2 \Lambda$, where $\Lambda$ is the negative cosmological constant in five dimensions, as expected.\\

The parameters $a$ and $b$ in $A_{1}(z)$ can be determined by comparing the holographic results with lattice QCD results at zero chemical potential. For example, as we will show shortly, the gravity solution with the above form of $A_{1}(z)$ undergoes a small to large black hole phase transition. This phase transition has been suggested to be dual to confinement-deconfinement phase transition in the boundary side \cite{He:2013qq,Yang:2015aia}. Demanding the critical temperature of this transition at zero chemical potential to be around $270 \ \text{MeV}$~\footnote{See \cite{Lucini:2003zr} for a lattice determination of the deconfinement temperature for pure gauge theories for several values of $N$. Only a mild dependence on (growing) $N$ is to be reported. }fixes the parameters $a$ and $b$ as
$$a=\frac{c}{9}, \ \ b=\frac{5 c}{16}.$$
We again like to emphasize that eq. (\ref{metsolution1}) is a solution of Einstein-Maxwell-dilaton system for any $A(z)$, and eq.~(\ref{Aansatz1}) is just a particular form. We chose this expression to reproduce some of the QCD results holographically, after which new predictions for other quantities can be made without further introduction of new parameters.

\subsection{Black hole thermodynamics}
Let us first discuss the thermodynamics of the gravity system with $A(z)$ as in eq.~(\ref{Aansatz1}). All the results in this section should be understood to be derived from this form of $A(z)$.\\
\begin{figure}[h!]
\begin{minipage}[b]{0.5\linewidth}
\centering
\includegraphics[width=2.8in,height=2.3in]{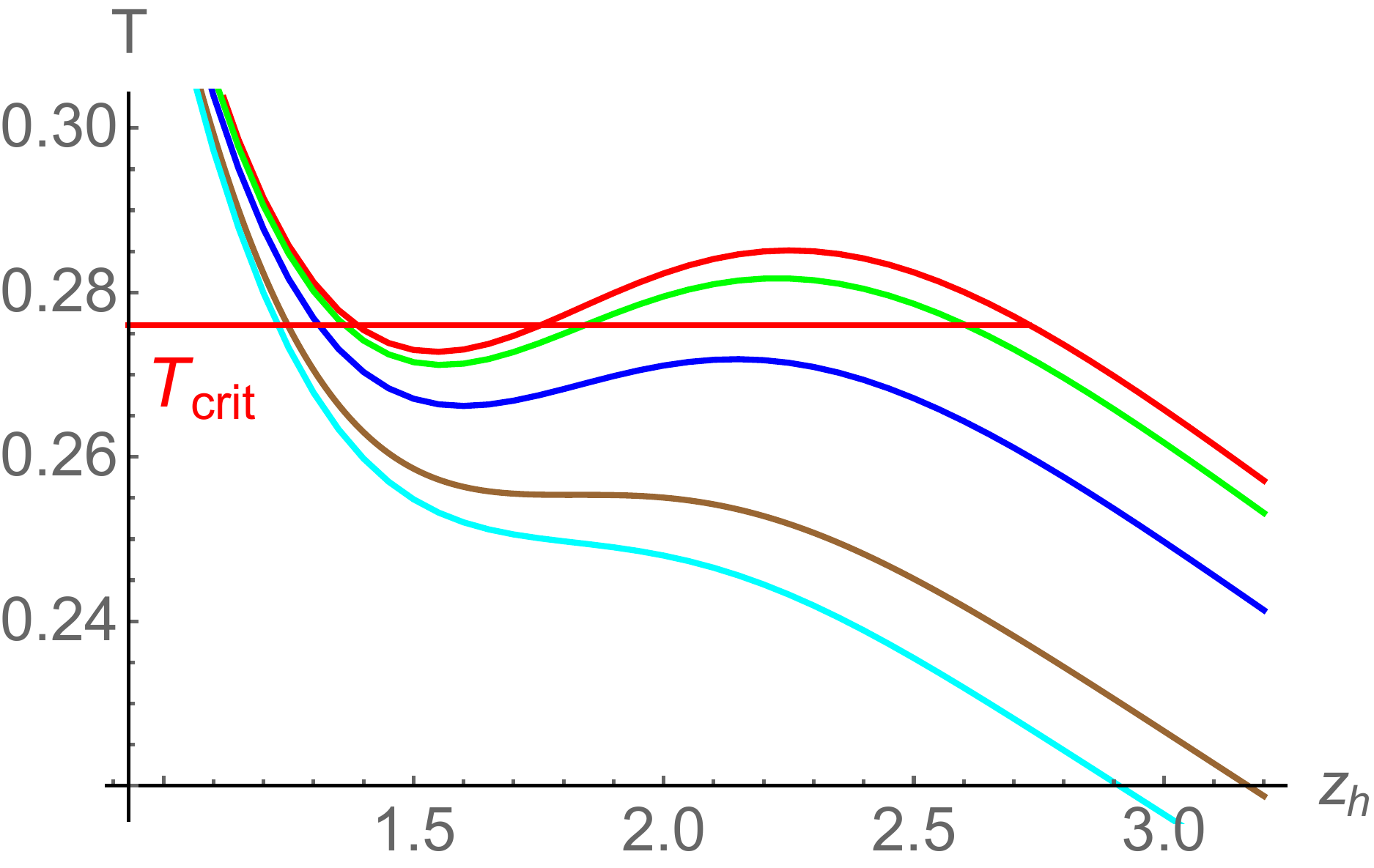}
\caption{ \small $T$ as a function of $z_h$ for various values of the chemical potential $\mu$. Here red, green, blue, brown and cyan curves correspond to $\mu=0$, $0.1$, $0.2$, $0.312$ and $0.35$ respectively. In units \text{GeV}.}
\label{zhvsTblackholecase1}
\end{minipage}
\hspace{0.4cm}
\begin{minipage}[b]{0.5\linewidth}
\centering
\includegraphics[width=2.8in,height=2.3in]{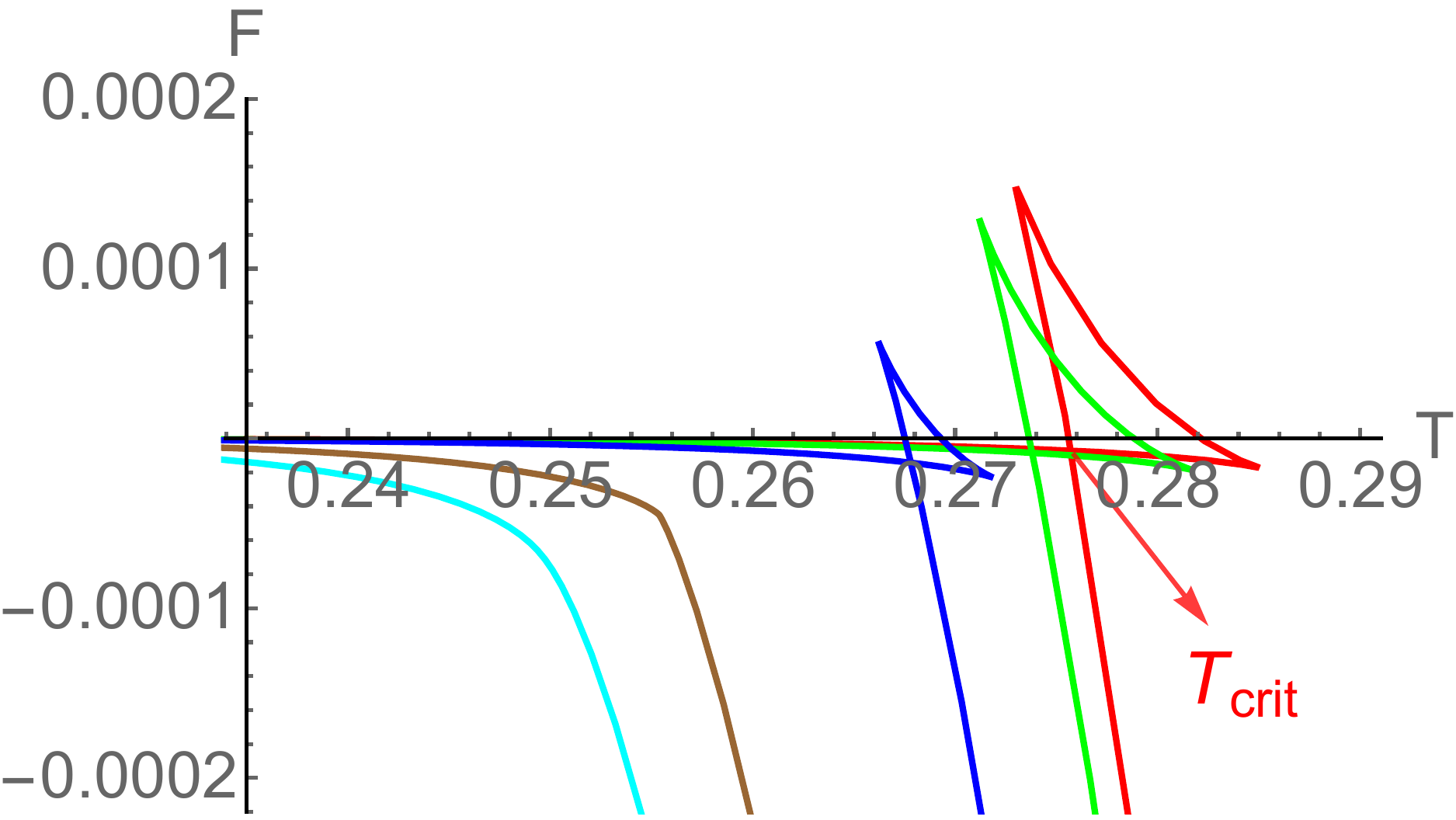}
\caption{\small $F$ as a function of $T$ for various values of the chemical potential $\mu$. Here red, green, blue, brown and cyan curves correspond to $\mu=0$, $0.1$, $0.2$, $0.312$ and $0.35$ respectively. In units \text{GeV}.}
\label{TvsFblackholecase1}
\end{minipage}
\end{figure}
\begin{figure}[h!]
\centering
\includegraphics[width=2.8in,height=2.3in]{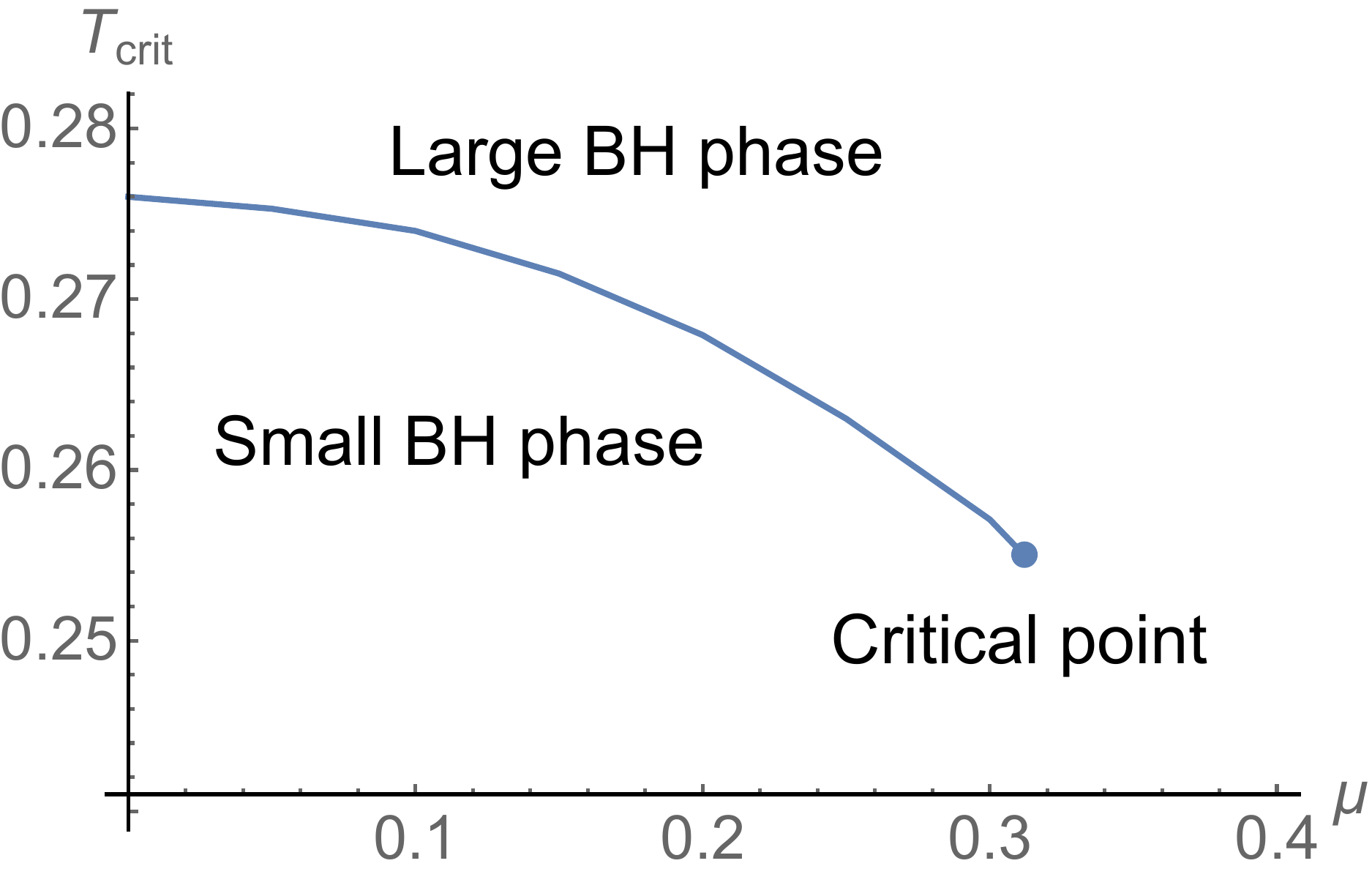}
\caption{ \small $T_{crit}$ as a function of $\mu$. For small $\mu$, there is a first order phase transition line between large and small black hole phases. This first order phase transition line terminates at a second order critical point. In units GeV.}
\label{MuvsTcrit}
\end{figure}

The variation of Hawking temperature with respect to horizon radius $z_h$ for various values of chemical potential is shown in Figure~\ref{zhvsTblackholecase1}. We find that for small chemical potential there are three branches in the $(T,z_h)$ plane, out of which one branch is stable, one metastable and the last branch is unstable. The stable and metastable branches, for which the slope in $(T,z_h)$ plane is negative, correspond to small and large black hole phases. For the unstable branch, the slope is positive. This suggest a phase transition form a small black hole (large $z_h$) to a large black hole (small $z_h$) as we steadily increase the Hawking temperature. This is indeed the case as can be observed from the free energy behaviour which is shown in Figure~\ref{TvsFblackholecase1}. In Figure~\ref{TvsFblackholecase1}, we used the normalization such that the free energy of thermal-AdS is zero. We see that the free energy has a swallowtail like structure --- a characteristic feature of a first order phase transition --- for small chemical potential. The unstable branch corresponds to the base of the swallowtail and a phase transition occurs at the kink. The corresponding temperature at the kink around which the free energy of the large black hole becomes larger than that of the small black hole defines the critical temperature $T_{crit}$. For $\mu=0$, we find $T_{crit}=0.276 \ \text{GeV}$.  Importantly, the free energy is always negative in stable branches which indicates that the black hole phase always has a lower free energy than thermal-AdS and hence is more stable.\\

However, the above pattern changes as we increase the value of $\mu$. For higher values of $\mu$, the size of the swallowtail starts to decrease and it completely disappears above a certain critical chemical potential $\mu_c$. For $A(z)=A_{1}(z)$, we find $\mu_c=0.312 \ \text{GeV}$. At $\mu_c$, the large and small black hole branches merge together and form a single stable black hole branch which is stable at all temperature. This can be seen from cyan curve for which $\mu=0.35 \ \text{GeV}$ is considered. The overall dependence of $T_{crit}$ on $\mu$ is shown in Figure~\ref{MuvsTcrit}. We find that $T_{crit}$ decreases with $\mu$. The solid line in Figure~\ref{MuvsTcrit} which separates the small and large black hole phases terminates at the second order critical point. This behaviour is analogous to the Van der Waal like liquid-gas phase transition and has been thoroughly discussed in context of charged AdS black holes in many cases \cite{Chamblin:1999tk,Chamblin:1999hg,Mahapatra:2016dae,Dey:2015ytd,Kubiznak:2012wp,Caldarelli:1999xj}.\\

The above small-large black hole phase transition on the gravity side has been suggested to correspond to the confinement-deconfinement phase transition in the dual boundary side in \cite{He:2013qq,Yang:2015aia}~\footnote{In \cite{He:2013qq,Yang:2015aia}, a different expression for $A(z)$ was used.}. In particular, the small black hole phase was suggested to be dual to confinement whereas the large black hole phase was suggested to be dual to deconfinement. However, as we will show shortly, this interpretation is not entirely correct. Especially, there are a few non-trivial issues in interpreting the small black hole phase as dual to confinement. Again by confinement we simply mean a phase for which the Wilson loop expectation value satisfies the area law while the expectation value of the Polyakov loop is zero. We will show in the next section that by taking the correct expression of the $q\overline q$ free energy neither of these conditions are \textit{strictly} satisfied in the boundary theory dual to the small black hole phase. However interestingly, despite of these limitations in interpreting the small black hole phase as being dual to genuine confinement, the entropy of the $q\overline q$ pair and speed of sound in this phase turn out to be in good qualitative agreement with lattice QCD in its confined phase.\\

\subsection{Free energy of the $q\overline q$ pair}
In order to study the Wilson and Polyakov loops, we first need to access the free energy of a $q\overline q$ pair which can be easily computed holographically from the world sheet on-shell action.\\

The gauge/gravity correspondence relates the free energy $\mathcal{F}$ of a $q\overline q$ pair to the on-shell action of a fundamental string in the dual gravity background. In particular, the $\mathcal{F}$ of a $q\overline q$ pair separated by a distance $\ell$ and evolving over time $t$ can be calculated by the on-shell action of the fundamental string, with the boundary condition that at the asymptotically AdS boundary, the string world sheet shares the rectangular boundary of sides $\ell$ and $T$. We have
\begin{eqnarray}
\mathcal{F}(T,\ell)=T \ S_{NG}^{on-shell}
\label{Sonshell}
\end{eqnarray}
where $T$ is the temperature and $S_{NG}^{on-shell}$ is the open string on-shell Nambu-Goto action,
\begin{eqnarray}
S_{NG}=\frac{1}{2 \pi \ell_{s}^2}\int d\tau d\sigma \sqrt{-\det g_s}, \ \ \ (g_s)_{\alpha\beta}=(g_s)_{MN}\partial_\alpha X^{M} \partial_\beta  X^{N}.
\label{NGaction}
\end{eqnarray}
Here, $g_s$ denotes the gravity background in string frame~\footnote{In the following, a subscript ``$s$" is used to denote quantities in the string frame.}, $X^{M}(\tau,\sigma)$ denote the open string embedding, $T_s=1/2 \pi \ell_{s}^2$ is the open string tension and $(\tau,\sigma)$ are the world sheet coordinates. Depending upon on the background geometry there can be multiple open string solutions and below we will show that each solution provides a unique result which can mimic the results of lattice QCD.\\

Here we will work in the static gauge with $\tau=t$ and $\sigma=y_1$. In this case there can be two world sheet configurations that minimize the Nambu-Goto action: a connected and a disconnected one. The connected world sheet is a $\cup$-shape configuration which extends from the boundary ($z=0$) into the bulk. In this case, we have the following expression for the free energy of the $q\overline q$ pair
\begin{eqnarray}
\mathcal{F}_{con}=\frac{L^2}{\pi \ell_{s}^2}\int_{0}^{z_*} dz \frac{z_{*}^2}{z^2} \frac{\sqrt{g(z)}e^{2A_{s}(z)-2A_{s}(z_*)}}{\sqrt{g(z)z_{*}^4 e^{-4A_{s}(z_*)} - g(z_*)z^4 e^{-4A_{s}(z)}}}
\label{Fcon}
\end{eqnarray}
where $z_*$ is the turning point of the connected world sheet. This turning point is related to the length of the $q\overline q$ pair as
\begin{eqnarray}
\ell=2 \int_{0}^{z_*} dz \ z^2\sqrt{\frac{g(z_*)}{g(z)}} \frac{e^{-2A_{s}(z)}}{\sqrt{g(z)z_{*}^4 e^{-4A_{s}(z_*)} - g(z_*)z^4 e^{-4A_{s}(z)}}}.
\label{lengthcon}
\end{eqnarray}
On the other hand, the disconnected configuration consists of two lines which are separated by distance $\ell$ and are extended from the boundary to the horizon,
\begin{eqnarray}
\mathcal{F}_{discon}=\frac{L^2}{\pi \ell_{s}^2}\int_{0}^{\bar{z}} dz \frac{e^{2 A_{s}(z)}}{z^2}.
\label{Fdiscon}
\end{eqnarray}
$\mathcal{F}_{discon}$ is independent of $z_*$ and therefore of the $q\overline q$ separation length $\ell$ as well. Here, $\bar{z}=z_h$ for the black background and $\bar{z}=\infty$ for thermal-AdS. It is important to mention that both $\mathcal{F}_{con}$ and $\mathcal{F}_{discon}$ are divergent quantities. The divergence arises from the $z=0$ part of the integral. In order to regularize the free energy, we will use the temperature independent renormalization scheme suggested in \cite{Ewerz:2016zsx}, which amounts to minimally subtracting the pole when cutting at the integral at $z=\varepsilon\ll 1$.  However, in most part of this paper we will deal with the difference in free energy anyhow where the diverging parts trivially cancel out.\\

\begin{figure}[t!]
\begin{minipage}[b]{0.5\linewidth}
\centering
\includegraphics[width=2.8in,height=2.3in]{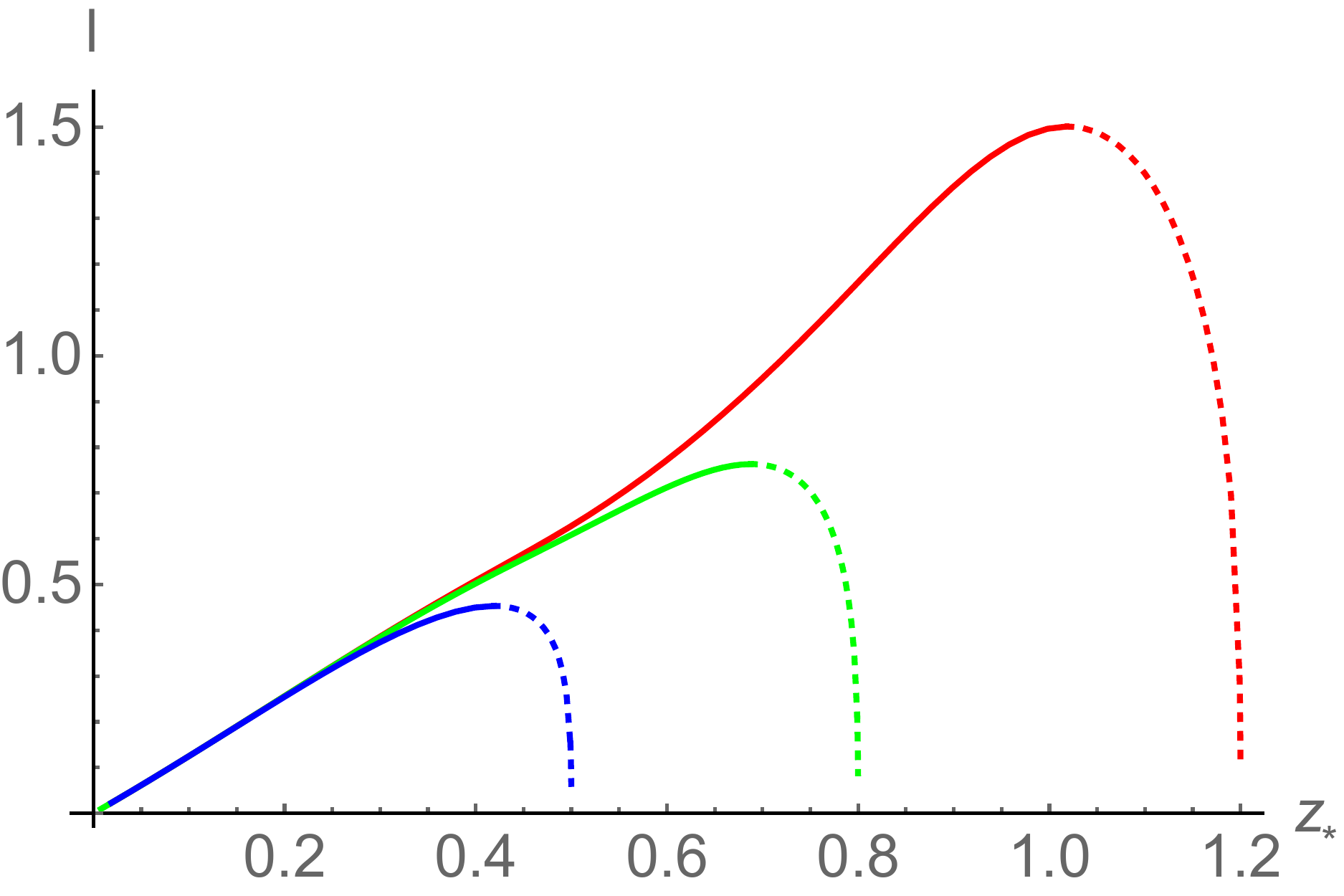}
\caption{ \small $\ell$ as a function of $z_*$ in the large black hole background for various values of $z_h$. Here $\mu=0$ and red, green and blue curves correspond to $z_h=1.2$, $0.8$ and $0.5$ respectively. In units \text{GeV}.}
\label{lvszsvszhMu0largecase1}
\end{minipage}
\hspace{0.4cm}
\begin{minipage}[b]{0.5\linewidth}
\centering
\includegraphics[width=2.8in,height=2.3in]{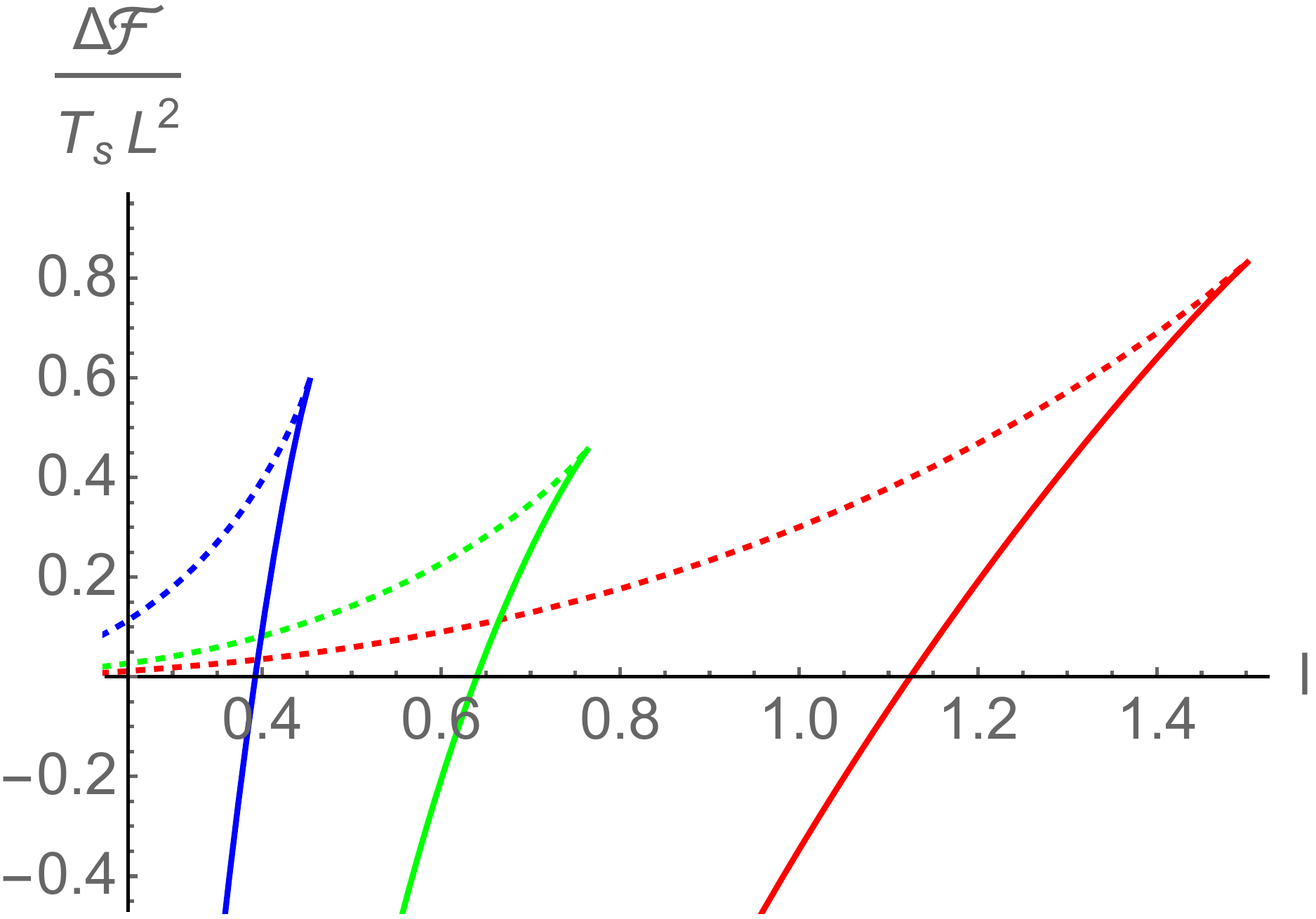}
\caption{\small $\Delta\mathcal{F}=\mathcal{F}_{con}-\mathcal{F}_{discon}$ as a function of $\ell$ in the large black hole background for various values of $z_h$. Here $\mu=0$ and red, green and blue curves correspond to $z_h=1.2$, $0.8$ and $0.5$ respectively. In units \text{GeV}.}
\label{lvsFvszhMu0largecase1}
\end{minipage}
\end{figure}

Let us first consider the large black hole phase (small $z_h$) as a gravity background. In Figure~\ref{lvszsvszhMu0largecase1}, the length $\ell$ as a function of $z_*$ for various values of the horizon radius is plotted. We observe that for every horizon radius there exists an $\ell_{max}$ above which the connected string configuration does not exist. We further find that there are two solutions for a given $\ell$: one for small $z_*$ (solid lines) and one for large $z_*$ (dotted lines). As we will see shortly, the one with smaller $z_*$ corresponds to an actual minimum of the string action whereas the one with larger $z_*$ corresponds to a saddle point.\\

In Figure~\ref{lvsFvszhMu0largecase1}, we have shown the difference in free energy between the connected and disconnected string solutions. In this Figure, solid and dotted lines correspond to smaller and larger branches of $\ell$ respectively. We observe that the former branch always has a lower free energy than the latter branch, indicating that it is a true minimum. However, we also observe that $\Delta\mathcal{F}$ can be greater or less than zero depending on the value of $\ell$. This suggests a phase transition from a connected to a disconnected string solution as we increase the (QCD) string length $\ell$. The string length at which $\Delta\mathcal{F}$ turns from negative to positive value defines the critical length $\ell_{crit}$. We find that $\ell_{crit}$ increases with increasing $z_h$. This suggests that for larger size black holes (small $z_h$), the connected string configuration confines closer to the AdS boundary. The behaviour that $\ell_{crit}$ decreases with temperature is consistent with the physical expectation that at higher and higher temperatures the boundary meson state would eventually melt to a free quark and antiquark (deconfined phase), a situation which is on the dual gravity side described by the disconnected string configuration. Since for large separations, this disconnected string configuration which is independent of separation length $\ell$ is more favourable, therefore the corresponding free energy of the $q\overline q$ pair is also independent of $\ell$. It implies that the QCD string tension is zero and that there is no linear law confinement in the boundary theory dual to the large black hole phase.\\
\begin{figure}[t!]
\begin{minipage}[b]{0.5\linewidth}
\centering
\includegraphics[width=2.8in,height=2.3in]{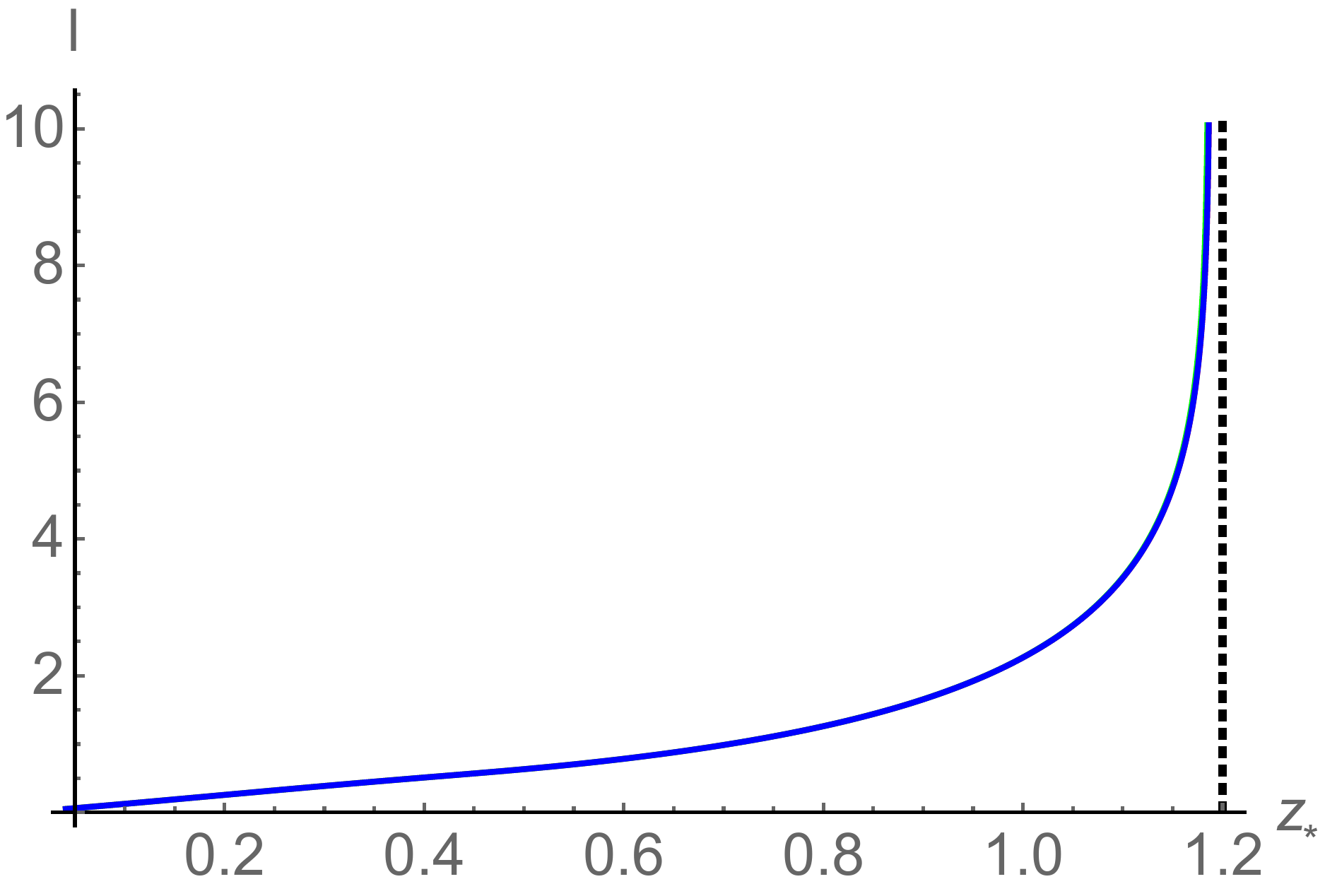}
\caption{ \small $\ell$ as a function of $z_*$ in the small black hole background for various values of $z_h$. Here $\mu=0$ and red, green and blue curves correspond to $z_h=3.0$, $3.5$ and $4.0$ respectively. In units \text{GeV}.}
\label{lvszsvszhMu0smallcase1}
\end{minipage}
\hspace{0.4cm}
\begin{minipage}[b]{0.5\linewidth}
\centering
\includegraphics[width=2.8in,height=2.3in]{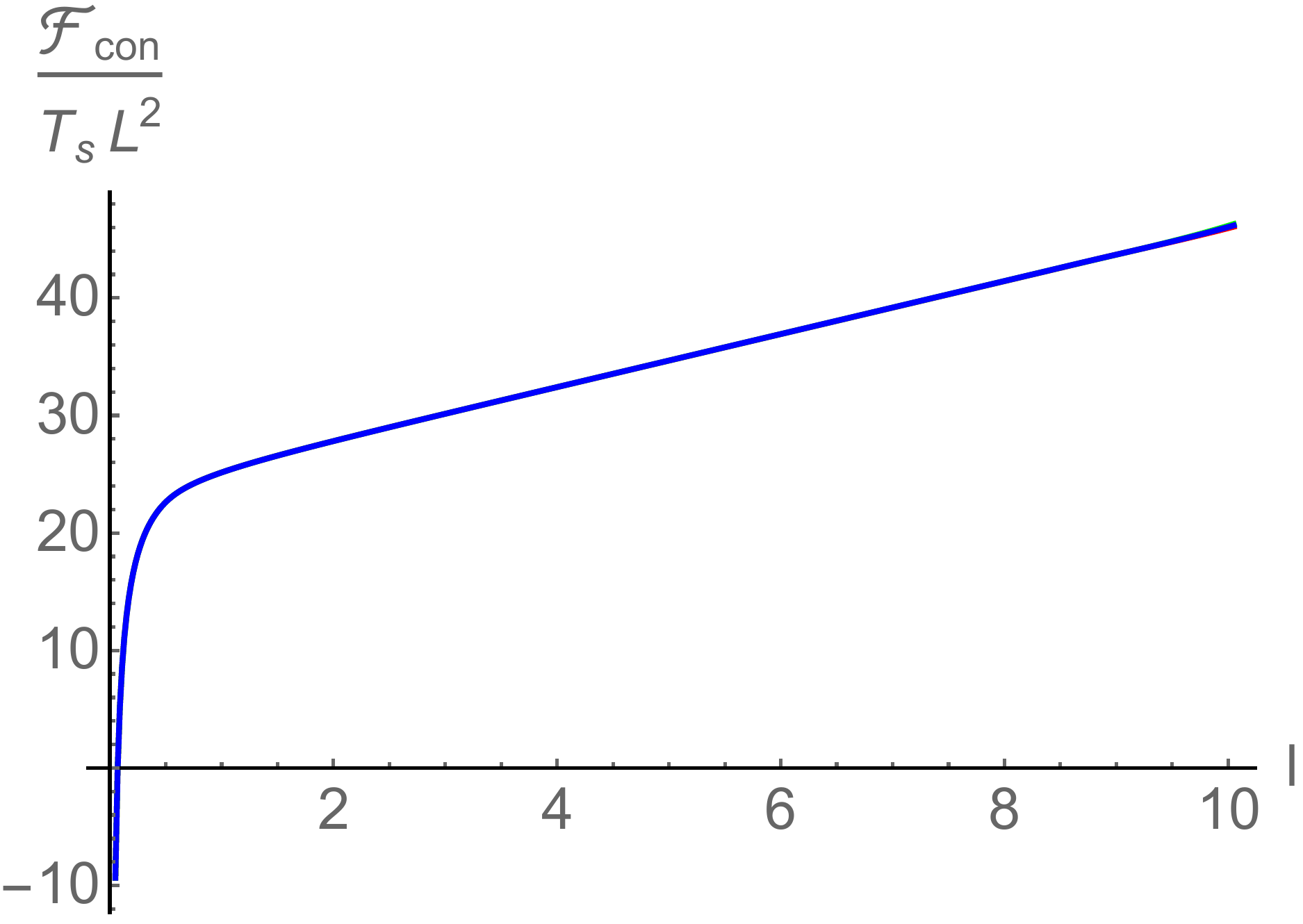}
\caption{\small $\mathcal{F}_{con}$ as a function of $\ell$ in the small black hole background for various values of $z_h$. Here $\mu=0$ and red, green and blue curves correspond to $z_h=3.0$, $3.5$ and $4.0$ respectively. In units \text{GeV}.}
\label{lvsFvszhMu0smallcase1}
\end{minipage}
\end{figure}

We now discuss the free energy of a $q\overline q$ pair in the small black hole phase (large $z_h$). We find that for a small black hole background, $\ell_{max}$ does not exist and $\ell$ continuously increases with $z_*$. This is shown in Figure~\ref{lvszsvszhMu0smallcase1}. The string world sheet does not penetrate deep into the bulk and saturates near $z\simeq1.2 \ \text{GeV}^{-1}$, suggesting some kind of an ``imaginary wall'' in the bulk AdS which cannot be penetrated by the string world sheet. Around this wall, $\ell$ also increases rapidly. At first sight this \textit{naively} suggests that in the small black hole background, the $q\overline q$ pair is always connected by an open string and forms a confined state. However, we need to be careful with this interpretation as we will show shortly. The regularized part of the corresponding free energy curve is shown in Figure~\ref{lvsFvszhMu0smallcase1}. We find that $\mathcal{F}_{con}$ is negative for small $\ell$ and increases linearly for large $\ell$. In fact, it can be shown  analytically that $\mathcal{F}_{con} \propto -1/\ell$  for small $\ell$ exhibiting a Coulomb potential, and  $\mathcal{F}_{con} = \sigma_s \ell$ for large $\ell$ suggesting confinement \cite{He:2013qq}.
Here $\sigma_s$ is the QCD string tension. Therefore, from the connected string configuration in the small black hole background we get the famous Cornell expression \cite{Eichten:1978tg}  $$\frac{\mathcal{F}_{con}}{T_s L^2}=-\frac{\kappa}{\ell}+\sigma_s \ell +\ldots$$ for the energy of a $q\overline q$ pair. By comparing the lattice QCD estimate of the string tension, $\sigma_s \approx 1/(2.34)^2 \text{GeV}^2$ \cite{Sommer:1993ce}, with our numerical results, we can further fix the value of the open string tension $T_s$ in units of the AdS length scale $L^2$. By doing that we find $T_s L^2\simeq 0.1$. This small value of $T_s L^2$, corresponding to a small 't Hooft coupling, generally indicates the breakdown of the classical gravity approximation and it suggests that the higher derivative $\alpha^\prime$ correction terms might become important. Although the small value of $T_s L^2$ is certainly a drawback of our model, however, we also like to point out that most of the phenomenological bottom-up gauge/gravity models suffer from this ambiguity (see for example \cite{Noronha:2009ud}, based on the self-consistent model of \cite{Gubser:2008yx,Gubser:2008ny} or \cite{Andreev:2006ct,Andreev:2006nw}, to name only a few.), and other models which do exhibit a large $T_s L^2$ generally have additional scaling symmetries in the dilaton potential (or additional parameters) which makes a large $T_s L^2$ arguable as it depends non-trivially on those additional scaling (as well as on the choice of dilaton normalization) \cite{Gursoy:2009jd}. See \cite{Kiritsis:2009hu} for further details on certain conceptual difficulties associated with the magnitude of $T_s L^2$ in bottom-up holographic models. Clearly it is an unsettled relevant question and more work is needed. Currently we do not have a resolution for this problem. Indeed, the size of $T_s L^2$ is only determined quantitatively upon selecting a ``stringy'' (loop related) QCD observable. In practice, this is usually achieved by matching on the string tension entering the Wilson loop, or, in the presence of dynamical quarks, the Polyakov-loop related free energy. If we refrain from matching to the precise QCD values, our qualitative findings stand as they are and are trustworthy. Keeping this ambiguity in mind, in the current work we will always express the free energy and entropy of the quark pair in (unspecified) units of $T_s L^2$. This is reasonable as our main aim here is to find a qualitative picture for the free energy and entropy of the quark pair from holography. Such approach is not uncommon, notice that also in \cite{Alho:2013hsa} thermodynamics of unquenched holographic QCD was discussed without actually fixing the string tension. Though, it must be kept in mind that quantitative holographic QCD loop-related results will only be trustworthy at the expense of having these quantities an order of magnitude smaller than expected from comparing with genuine QCD.\\

It is clear from the linear dependence of $\mathcal{F}_{con}$ on the inter-quark separation length that the Wilson loop of the dual boundary theory exhibits an area law, as one expects for a confined gauge theory, and that the boundary theory dual to the small black hole phase seems to satisfy the typical properties of confinement. This behaviour of $\mathcal{F}_{con}$  was used in \cite{He:2013qq,Yang:2015aia}  to motivate that the small black hole phase is actually dual to confinement.\\

However, as mentioned before, we need to be careful with the above interpretation. In particular, the connected string is not the only world sheet solution of the Nambu-Goto action and one needs to compare with the free energy of the disconnected string as well. Generally, in the gauge/gravity correspondence, the confined phase is dual to a thermal-AdS space. In that case the upper limit in the integral of $\mathcal{F}_{discon}$ is at $\bar{z}=\infty$, which makes $\mathcal{F}_{discon}$ divergent. Therefore, in that case the connected string always gives a lower free energy than the disconnected one. Consequently, one gets a linear law confinement for the $q\overline q$ pair from the connected string solution using the dual thermal-AdS background. However, with a black hole background, the disconnected strings extend from the asymptotic boundary to maximally the horizon at $\bar{z}=z_h$, which makes $\mathcal{F}_{discon}$ finite (of course after removing the usual UV divergences), and hence its free energy could become less than $\mathcal{F}_{con}$ as well. For this reason, it is now important to compare $\mathcal{F}_{con}$ and $\mathcal{F}_{discon}$ when one uses black hole backgrounds (like we did in the large black hole phase in Figure~\ref{lvsFvszhMu0largecase1}) to get the true minimum of Nambu-Goto action and correspondingly to study the properties of confinement and deconfinement.\\
\begin{figure}[t!]
\begin{minipage}[b]{0.5\linewidth}
\centering
\includegraphics[width=2.8in,height=2.3in]{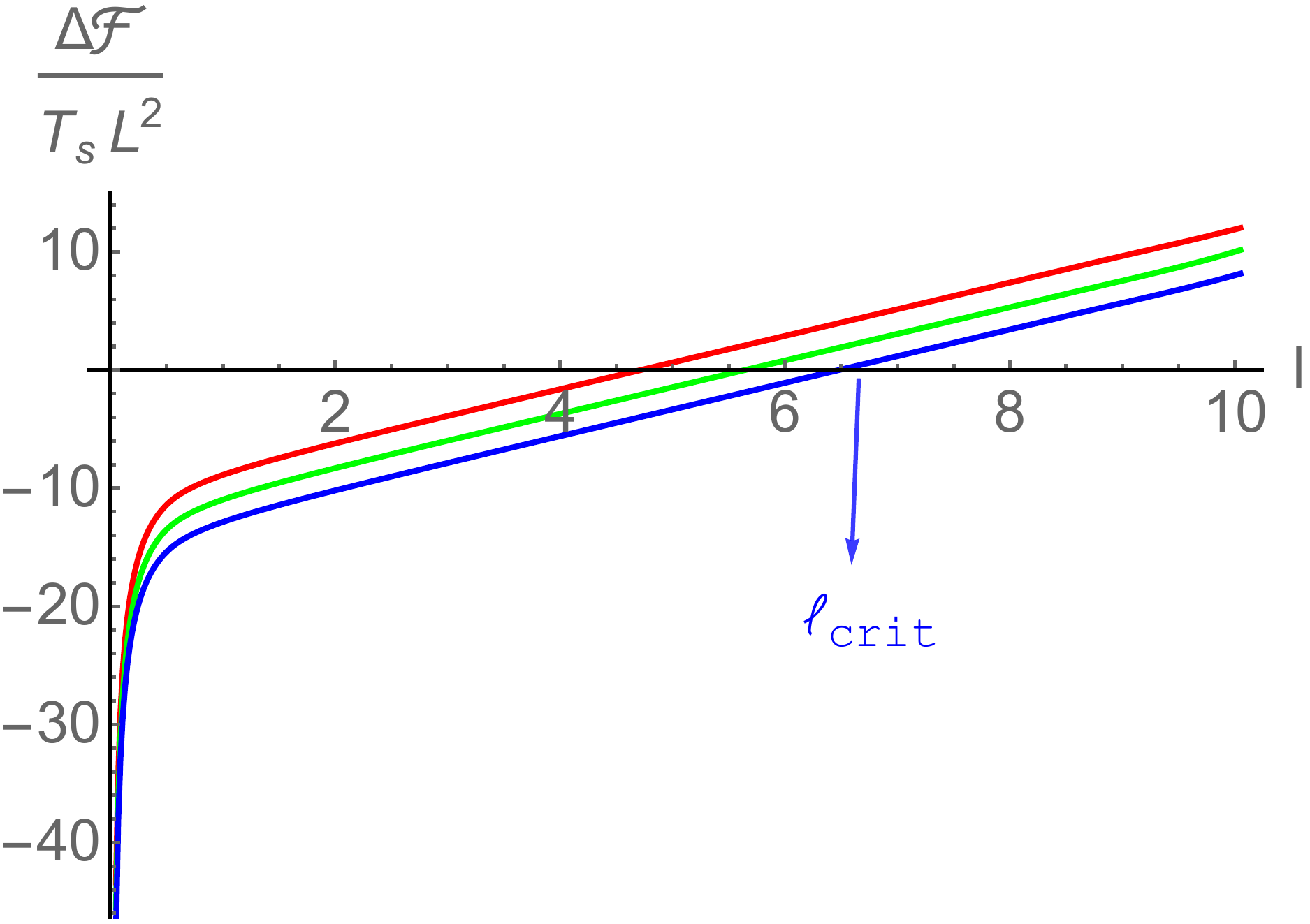}
\caption{ \small $\Delta\mathcal{F}=\mathcal{F}_{con}-\mathcal{F}_{discon}$ as a function of $\ell$ in the small black hole background for various values of $z_h$. Here $\mu=0$ and red, green and blue curves correspond to $z_h=3.0$, $3.5$ and $4.0$ respectively. In units \text{GeV}.}
\label{lvsdelFvszhMu0smallcase1}
\end{minipage}
\hspace{0.4cm}
\begin{minipage}[b]{0.5\linewidth}
\centering
\includegraphics[width=2.8in,height=2.3in]{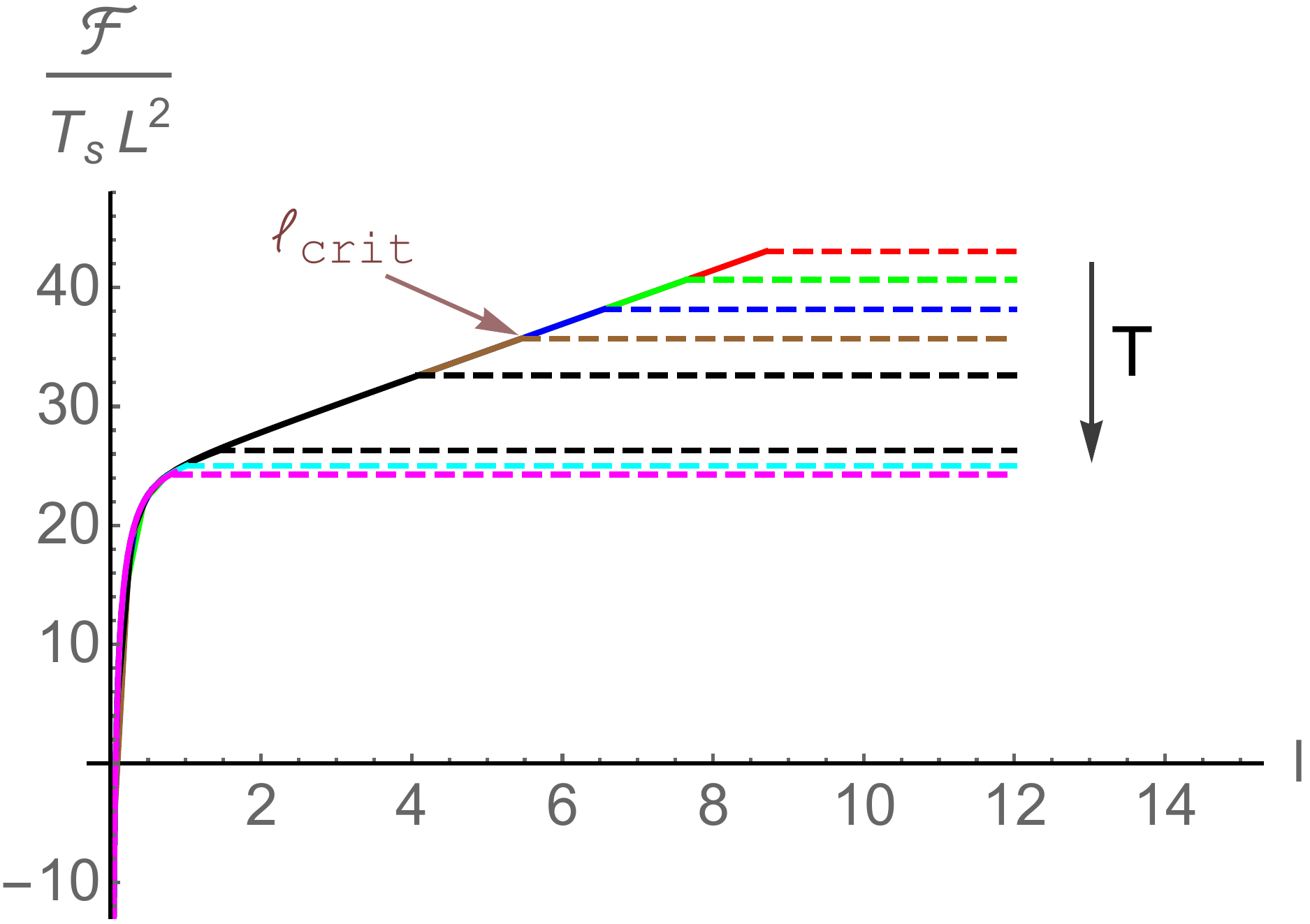}
\caption{\small $\mathcal{F}$ as a function of $\ell$ for various values of $T$. Here $\mu=0$ and red, green, blue, brown, black, cyan and magenta curves correspond to $T/T_{crit}=0.6$, $0.7$, $0.8$, $0.9$, $1.0$, $1.1$ and $1.2$ respectively. The direction of growing $T$ has been indicated. In units \text{GeV}. }
\label{lvsFvsTcase1}
\end{minipage}
\end{figure}

The comparison between $\mathcal{F}_{con}$ and $\mathcal{F}_{discon}$ in the small black hole background is shown in Figure~\ref{lvsdelFvszhMu0smallcase1}, where $\Delta\mathcal{F}=\mathcal{F}_{con}-\mathcal{F}_{discon}$ is plotted as a function of $\ell$  for the same parameter values as in Figure~\ref{lvsFvszhMu0smallcase1}. We see that $\Delta\mathcal{F}$ can indeed be greater than zero, especially for large $q\overline q$ separation length. This implies that $\mathcal{F}_{discon}$ is actually the true minimum for the string action for large $\ell$, and therefore, one should consider $\mathcal{F}_{discon}$ to study the (correct) properties of the $q\overline q$ pair. Accordingly, since $\mathcal{F}_{discon}$ is independent of $\ell$, the string tension is zero and hence there is no more linear law confinement in the boundary theory dual to small black hole phase. Therefore, although $\mathcal{F}_{con}$ does indeed show an area law for the Wilson loop, however, it is $\mathcal{F}_{discon}$ that is relevant for large $\ell$, not leading to an area law. Further details are presented in Figure~\ref{lvsFvsTcase1}, where flattening of inter-quark energy at large $\ell$ is explicitly shown.\\

In Figure~\ref{lvsFvsTcase1}, the behaviour of the $q\overline q$ free energy as a function of $\ell$ for various temperatures is shown. For $T<T_{crit}$ (lines above the upper black line), we are in the small black hole phase; and for $T>T_{crit}$, (lines below the 2nd black line), we are in the large black hole phase. The two black lines correspond to $T=T_{crit}$ at which the small and the large black hole phases (having different horizon radius) coexist. We find that the string tension for $T\leq T_{crit}$, calculated from the slope of the linear part of $\mathcal{F}$, is almost temperature-independent and remains non-zero even at $T=T_{crit}$, a result supported by lattice QCD as well \cite{Kaczmarek:1999mm,Cardoso:2011hh}. Moreover, for $T > T_{crit}$ the size of the linear part in $\mathcal{F}$ starts to decrease and so is the string tension. However, as opposed to its lattice QCD counterpart, there exists a small discontinuity in the string tension at $T=T_{crit}$. This can again be traced back to the first order phase transition between the small and large black hole phases at $T=T_{crit}$, where both phases coexist. \\

For completeness, let us notice that according to Figure~\ref{lvsFvsTcase1}, the critical length $\ell_{crit}$ at which the string tension vanishes in the \emph{specious-confined} phase, decreases with increasing temperature. A similar signal is obtained from the entanglement entropy using the same model \cite{wip}. \\

To appreciate the length scales $\ell_{crit}$ in Figure~\ref{lvsFvsTcase1}, let us remind here the typical scale of heavy quark bound states, like charmonia, having a binding size of the order of $0.5~\text{fm} \simeq 2.5~\text{GeV}^{-1}$ \cite{Satz:2005hx}. In \cite{Cardoso:2011hh}, the linear potential at $T=0$ was considered up to $\ell \simeq 4 \sigma_0 \simeq 9~ \text{GeV}^{-1}$.~\footnote{$\sigma_{0}$ is the string tension at $T=0$.} This strongly suggests that the values of $\ell_{crit}$ obtained here for temperatures not too far from $T_{crit}$, are well compatible with QCD phenomenologically relevant length scales.\\

We further like to add that these results are valid irrespective of the value of the chemical potential. For example, for finite $\mu<\mu_c$ we again find $\Delta\mathcal{F}>0$ for large $\ell$. Moreover, at a fixed temperature, $\ell_{crit}$ is found to decrease with $\mu$ in the small black hole phase.\\

Further evidence for the problem of correctly interpreting the small black hole phase as the gravity dual of confined phase can be inferred from the Polyakov loop expectation value. As said before, and well known in the literature, the Polyakov loop acts as an order parameter for the confinement/deconfinement phase transition, vanishing in the confined phase and attaining a non-zero value in the deconfined phase. Therefore it is of importance to find out how the expectation value of Polyakov loop behaves in the current model.\\

The expectation value of the Polyakov loop can be calculated holographically from the heavy quark free energy. The regularized form of the latter (after removing the UV divergence) is given by
\begin{eqnarray}
\mathbf{F_{reg}}=\frac{\mathcal{F}_{discon}}{2},
\label{Polyakov}
\end{eqnarray}
from which one can extract the Polyakov loop $$\mathbf{P}=e^{-\mathbf{F_{reg}}/T}.$$
We see that $\mathbf{P}$ can be zero in confined phase when either $T$ is zero or $\mathbf{F_{reg}}$ is divergent. In the small black hole phase, none of these conditions is met. For example, $T$ is finite for the small black hole phase and, as discussed above, $\mathbf{F_{reg}}$ does not have an IR divergence. Therefore, the Polyakov loop expectation value is \textit{strictly} non-zero for the boundary theory dual to the small black hole phase and consequently this dual boundary phase is not exactly corresponding to a confined phase.\\

We have followed the holograpic literature here and considered a single Polyakov loop, even in the presence of a non-vanishing chemical potential \cite{Cai:2012xh,Alho:2013hsa,Stoffers:2010sp}. We should however notice that the latter breaks the charge conjugation symmetry {\sf C} between the (unequal number of) quarks and anti-quarks, and as such the Polyakov loop expectation values for an isolated quark or anti-quark are not supposed to be equal \cite{Svetitsky:1985ye,Dumitru:2005ng,Reinosa:2015oua}. In the current approximation of the gauge-gravity duality to access loop quantities as the Polyakov loop, we are however unable to probe this {\sf C} breaking since the Nambu-Goto action is blind to ${\sf C}$. It would require to take into account stringy corrections to the classical Nambu-Goto worldsheet action (including back reaction corrections related to introducing either a quark or antiquark located at the end point of the string considered dual to the loop), a task far beyond this and most --not to say all-- other works on holographic QCD models.\\

We also like to point out another important issue here. Although, $\mathbf{P}$ is non-zero in the small black hole phase in a strict sense, however it is extremely small. For instance, $\mathbf{P}\sim 10^{-30}$ at $T= 0.9 \ T_c$, and its magnitude is even smaller at lower temperatures. Moreover, it is also clear from Figure~\ref{lvsdelFvszhMu0smallcase1} that the value of $\ell_{crit}$ at which $\Delta\mathcal{F}$ changes its sign, increases with decreasing temperature (or as $z_h$ increases). This suggest that for smaller and smaller temperature, the quark and antiquark form a bound state up to larger distances. For example, $\ell_{crit}$ at $T=0.5 \ T_c$ is an order of magnitude larger than at $T=1.5 \ T_c$. For these reasons, keeping in mind various technical issues in interpreting the small black hole phase as the gravity dual of the confined phase, we may call this dual phase the \textit{specious-confined} phase,  to distinguish it from the ``standard'' confined phase. We will see in the following sections that the entropy of the $q\overline q$ pair as well as the speed of sound in this \textit{specious-confined} phase is also rather similar to that of the lattice QCD confined phase.\\

Strangely, our \textit{specious-confined} phase resembles the behaviour of full (unquenched) QCD at the level of non-local observables such as Wilson or Polyakov loops expectation values, while at the level of thermodynamics, it becomes closer to quenched QCD. In full QCD, the Polyakov loop is not an exact order parameter because of the explicit breaking of the $\mathbb{Z}_N$ symmetry due to the dynamical quarks in the fundamental representation. Moreover, quark pair creation prevents the linear behaviour of the confining potential in full QCD, rather one observes a flattening (the so-called ``string breaking'') at larger values of $\ell$ \cite{Bali:2005fu}, not unlike what we find here. Indeed, as shown in Figure \ref{lvsFvsTcase1}, one can clearly observe the flattening of inter-quark energy for large $\ell$. Moreover, just like in full (unquenched) QCD, the inter-quark energy in our holographic model also approaches a temperature dependent constant value at larger values of $\ell$. This temperature dependent constant value decreases with temperature (from \textit{specious-confined} to deconfined), which again is in line with full QCD results (see for example Figure 1 of \cite{Kaczmarek:2005zp}). Although we do not have a clear understanding where this peculiar ``mixture'' of quenched vs.~unquenched QCD-like behaviour is caused by precisely, it is not related to having a too small 't Hooft coupling (or small $T_s L^2$) that would cloud the dual gauge-gravity interpretation: as we noticed before, the qualitative features of our models persist, independent on how one ultimately decides to choose by hand or fix on an external QCD variable the string tension, or similar quantity. \\Another possibility might be that it is related to a wrongful identification of the scalar field $\phi$ with the dilaton, see discussion below eq.~\eqref{stringmetric}, but as discussed there, not making this identification would lead to a loss of all confining-related properties of our models.

\subsection{Thermal entropy of a $q\overline q$ pair}
We now move on to discuss the thermal entropy $S$ of a $q\overline q$ pair in the above \textit{specious-confined}/deconfined phases. This entropy can be calculated from the $q\overline q$ free energy $\mathcal{F}$ via the relation,
\begin{eqnarray}
S=-\frac{\partial\mathcal{F}}{\partial T}.
\label{entropy}
\end{eqnarray}
For black hole backgrounds, we have two choices for $S$, corresponding to the two different behaviours of $\mathcal{F}$ with respect to the $q\overline q$ separation length. For large separation, we have
\begin{eqnarray}
S_{discon}(\ell>\ell_{crit})=-\frac{\partial\mathcal{F}_{discon}}{\partial T}.
\label{entropyBHlarge1}
\end{eqnarray}
On the other hand for small separation, we have
\begin{eqnarray}
S_{con}(\ell<\ell_{crit})=-\frac{\partial\mathcal{F}_{con}}{\partial T}.
\label{entropyBHlarge2}
\end{eqnarray}
\\
We will see that the two distinct behaviours of $\mathcal{F}$ as a function of the $q\overline q$ separation length precisely capture the QCD results for the entropy in their respective regimes.\\

\begin{figure}[h!]
\begin{minipage}[b]{0.5\linewidth}
\centering
\includegraphics[width=2.8in,height=2.3in]{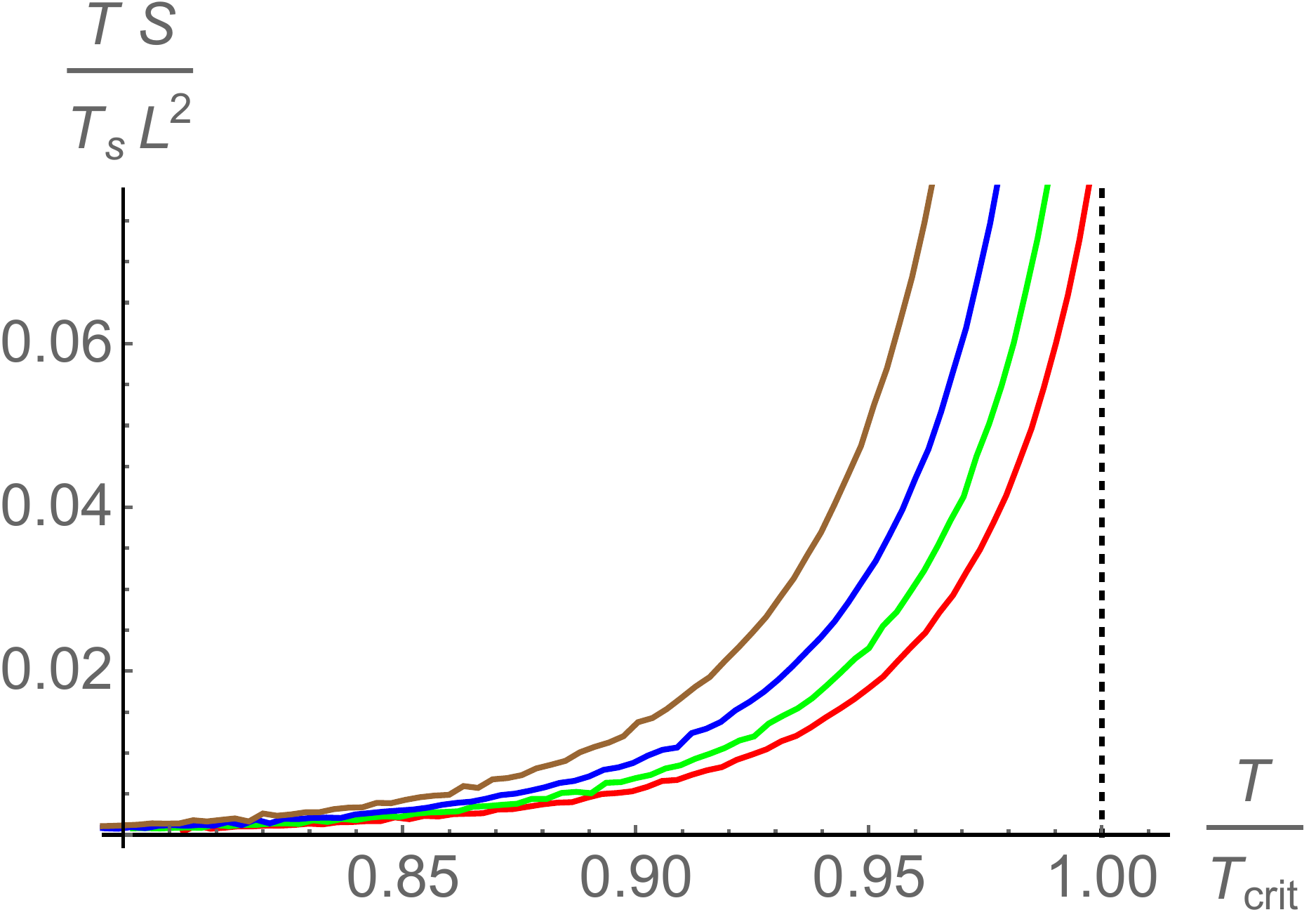}
\caption{ \small Entropy of the $q\overline q$ pair as a function of temperature in the \textit{specious-confined} phase for various values of the chemical potential $\mu$. Here red, green, blue and brown curves correspond to $\mu=0.10$, $0.15$, $0.20$ and $0.25$ respectively. In units \text{GeV}.}
\label{TvsSvsMuconfdcase1}
\end{minipage}
\hspace{0.4cm}
\begin{minipage}[b]{0.5\linewidth}
\centering
\includegraphics[width=2.8in,height=2.3in]{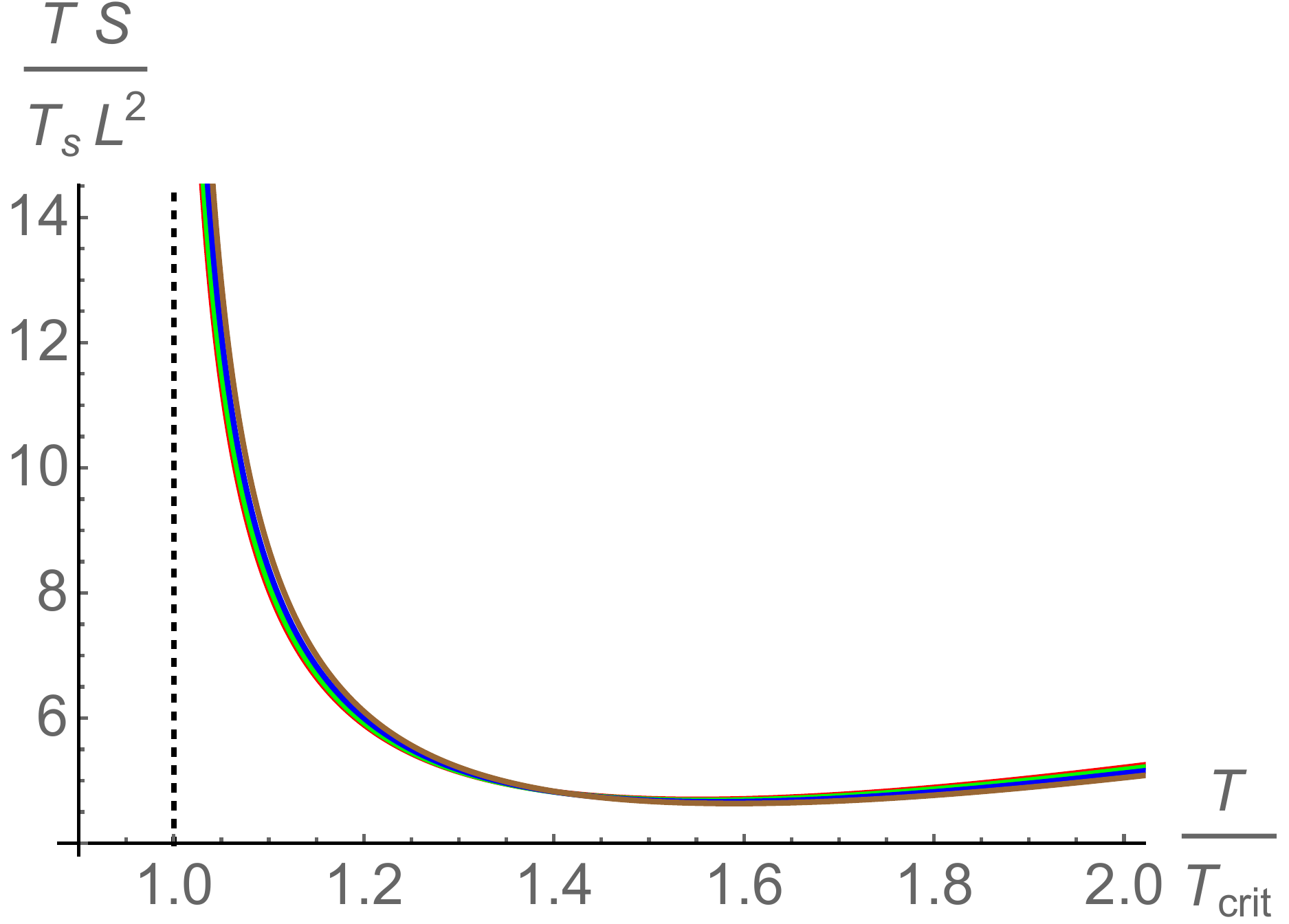}
\caption{\small Entropy of the $q\overline q$ pair as a function of temperature in the deconfined phase for various values of the chemical potential $\mu$. Here red, green, blue and brown curves correspond to $\mu=0.10$, $0.15$, $0.20$ and $0.25$ respectively. In units \text{GeV}.}
\label{TvsSvsMudeconfdcase1}
\end{minipage}
\end{figure}
We now discuss a few silent features of the holographic calculations for the $q\overline q$ entropy and compare with the lattice QCD results:

\begin{itemize}
 \item The variation of $S_{con}$ with respect to temperature in the small black hole phase is shown in Figure~\ref{TvsSvsMuconfdcase1}. Here, we have fixed the $q\overline q$ separation length to $\ell=2 \ \text{GeV}^{-1}$. The essential features of our analysis remain unchanged for other values of $\ell$, and $\ell=2 \ \text{GeV}^{-1}$ is just a particular choice. We find that in this \textit{specious-confined} phase the entropy initially varies slowly at low temperatures, and then increases rapidly towards the critical temperature. It indicates a large amount of entropy associated with the $q\overline q$ pair near the critical temperature, as also observed in lattice QCD. We see that the holographic results for the $q\overline q$ entropy in this \textit{specious-confined} phase are qualitatively similar to those of lattice QCD. It also indicates the main difference compared to \cite{Iatrakis:2015sua}, where the entropy in the confined phase was reported to be zero. We again like to point out that the non-zero entropy in the \textit{specious-confined} phase here arises precisely due to the fact that the dual gravity background is a (small) black hole, which depends on temperature. In the usual AdS/CFT correspondence, the confined phase is generally dual to thermal-AdS (without horizon and temperature) and therefore the entropy of quark pair is inherently zero in those confined phases.

     \item The entropy of the $q\overline q$ pair as a function of temperature in the large black hole phase at large separation is shown in Figure~\ref{TvsSvsMudeconfdcase1}. For this purpose we use eq.~(\ref{entropyBHlarge1}) for the entropy. We observe that similarly to the \textit{specious-confined} phase a large amount of entropy is associated with the $q\overline q$ pair near the critical temperature. Similar results in the deconfined phase were found in \cite{Iatrakis:2015sua} using the improved holographic model of \cite{GursoyI,GursoyII}~\footnote{The temperature dependence of the entropy of a $q\overline q$ pair and of the string tension were also discussed in a dual phenomenological model in \cite{Andreev:2006nw}. Unlike for our current model, the metric of \cite{Andreev:2006nw} was chosen by hand and it displays no clear phase transition picture on the gravity side. Rather it relied on an effective potential approach.}. We further find that the higher temperature asymptotics of the $q\overline q$ entropy in our model is similar to those of \cite{Iatrakis:2015sua}. In particular, for $T \gtrsim 2 T_c$, we find a tendency of a rise of $T S$ with temperature as also observed in lattice QCD. Moreover, our analysis further predicts similar asymptotic behavior of $T S$ in the presence of chemical potential too. \\
\begin{figure}[h!]
\begin{minipage}[b]{0.5\linewidth}
\centering
\includegraphics[width=2.8in,height=2.3in]{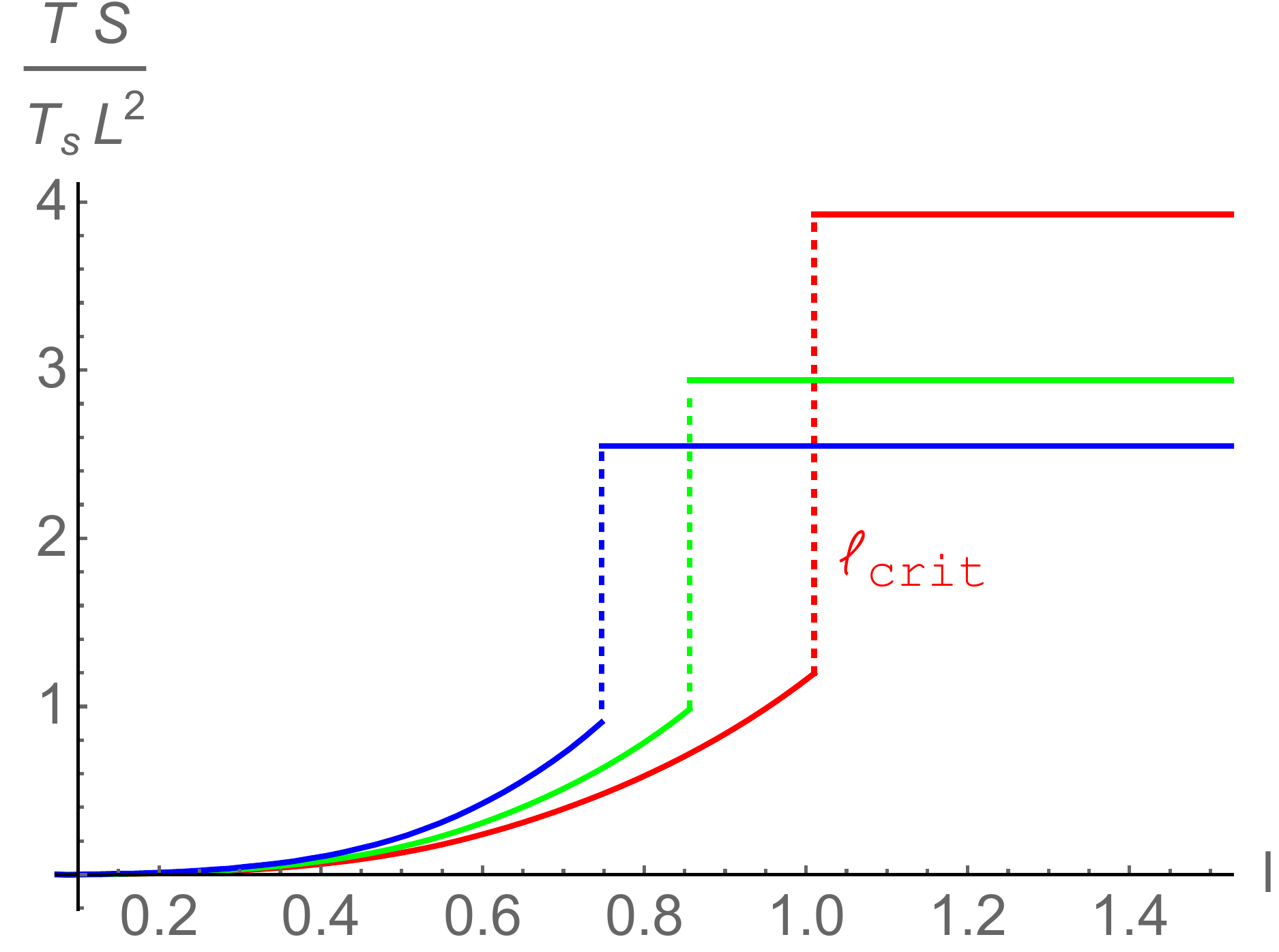}
\caption{ \small Entropy of the $q\overline q$ pair as a function of distance in the deconfined phase for various temperatures. Here $\mu=0$ and red, green and blue curves correspond to $T/T_{crit}=1.1$, $1.2$ and $1.3$ respectively. In units \text{GeV}.}
\label{lvsSvsTMu0case1}
\end{minipage}
\hspace{0.4cm}
\begin{minipage}[b]{0.5\linewidth}
\centering
\includegraphics[width=2.8in,height=2.3in]{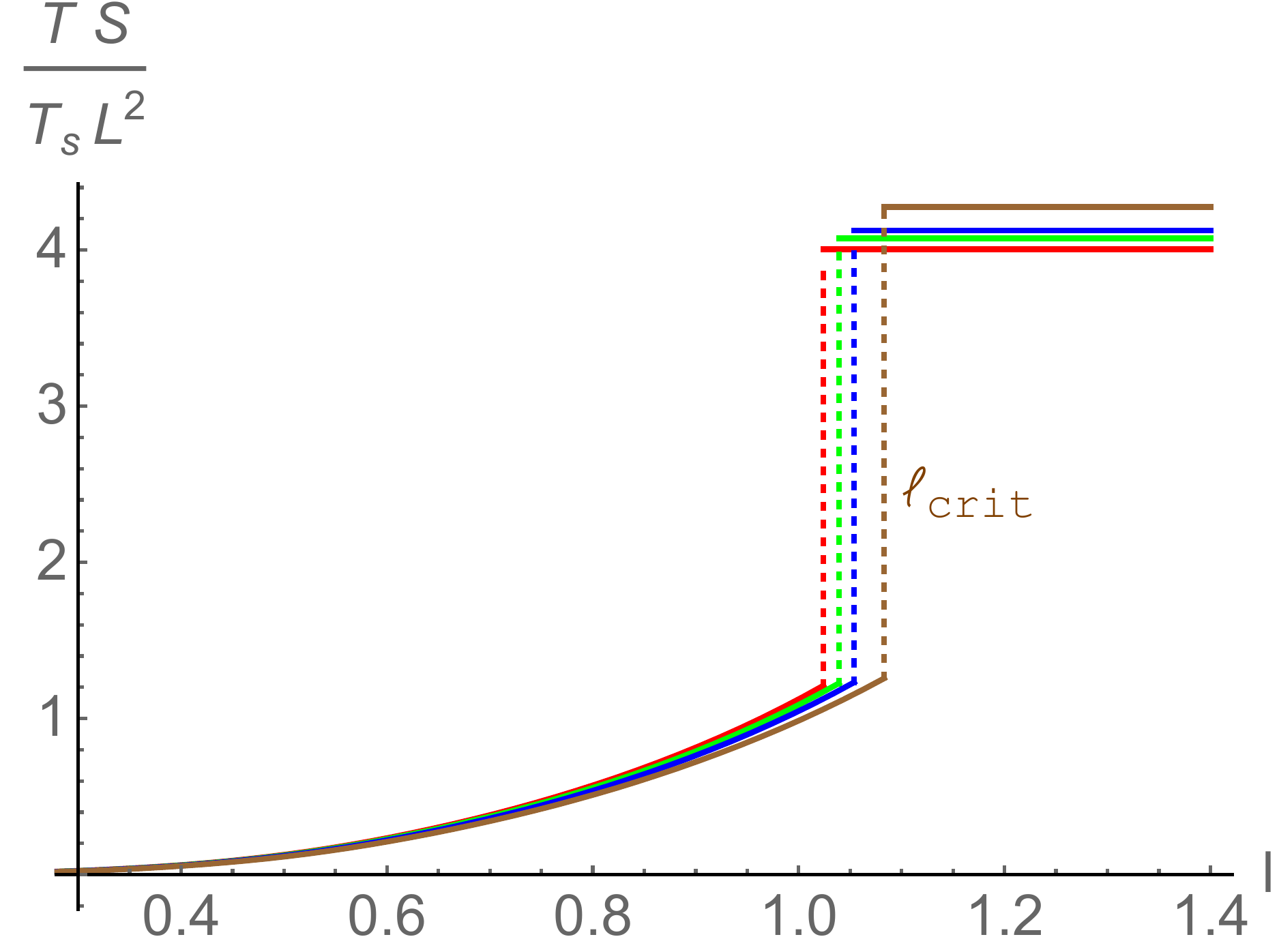}
\caption{\small Entropy of the $q\overline q$ pair as a function of distance in the deconfined phase for various chemical potentials. Here $T=1.1 T_{crit}$ and red, green, blue and brown curves correspond to $\mu=0.10$, $0.15$, $0.20$ and $0.25$ respectively. In units \text{GeV}.}
\label{lvsSvsMuwithT1Pt1case1}
\end{minipage}
\end{figure}
\item Another important lattice QCD result which the holographic model considered here correctly describes, is the increase in the entropy of the $q\overline q$ pair as a function of distance between them, see Figure~\ref{lvsSvsTMu0case1}. Here we have shown results in deconfined phase for three different temperatures. We see that for each case, $S$ increases with $\ell$. Moreover, for large $\ell$, $S$ saturates to a constant value and becomes independent of it~\footnote{In order to make it more readable, the magnitude of $S$ has been suppressed by a factor of 2 in $\ell>\ell_{crit}$ region of Figure~\ref{lvsSvsTMu0case1}.}. This is due to the fact that the disconnected string configuration has lower free energy at large separations, while it is independent of $\ell$. Therefore, the corresponding entropy is also independent of $\ell$. We see that these results qualitatively match with the results predicted by lattice QCD (shown in Figure~\ref{latticeQCD2}). However, as opposed to the latter, the entropy here does not smoothly go to saturation. There is a discontinuity in the entropy at $\ell_{crit}$ (denoted by dotted lines in Figure~\ref{lvsSvsTMu0case1}). This discontinuity in the entropy arises precisely due to the first order transition between the different string configurations at $\ell_{crit}$.

 \item In Figure~\ref{lvsSvsMuwithT1Pt1case1}, the variation of the entropy as a function of the inter-quark distance for various values of chemical potential is shown. We do not have lattice results here to compare with. In this regard, these entropic results can be thought of as a original prediction from holography. We find that essential features of the entropy remain the same even with a chemical potential present. The entropy again saturates to a constant value at large separations. Also, $\ell_{crit}$ increases with chemical potential. This suggest that the distance around which entropy  saturates, increases with the chemical potential.
\end{itemize}
We end this section by summarising our main results obtained so far. Till now we showed that by considering the form of $A(z)$ as in eq.~(\ref{Aansatz1}), a small/large black hole phase transition in the gravity side appears which on the dual boundary side corresponds to a \textit{specious-confined}/deconfined phase. We showed that although there is always a $\cup$-shape connected string configuration in the \textit{specious-confined} phase, it is however the disconnected string configuration which is more favourable for large $q\overline q$ separation length. Interestingly, we find that in spite of the reported differences between the \textit{specious-confined} and standard confined phase, the entropy of the quark pair in the former phase is qualitatively similar to lattice QCD confined phase results. Similarly, in the large black hole phase when the temperature is sufficiently high, it is the disconnected string configuration which is more favourable at larger $q\overline q$ separation. The two disconnected strings describe the dissociation of a bound state into a quark and antiquark, thereby corresponding to the deconfined phase in the dual boundary theory. The small/large black hole phases therefore correspond to \textit{specious-confinement}/deconfinement phases, and the corresponding critical temperature is found to decrease with chemical potential. Moreover, the entropy of the $q\overline q$ pair in this phase is in qualitative agreement with lattice QCD results.\\

\section{Case 2: standard confinement}
It is instructive to also study the entropy by considering a different form of $A(z)$. In particular, in order to check the universal nature of the results presented for the $q\overline q$ entropy, especially for the deconfined phase, we will now study the entropy in a different holographic model where the gravity dual genuinely describes the confinement/deconfinement phases in the boundary theory. For this purpose we need to consider a different gravity solution. As mentioned earlier, the same can be achieved by considering different form of $A(z)$.\\

Let us now consider another simple form of $A(z)$, such as
\begin{eqnarray}
A(z)=A_{2}(z)=- \bar{a} z^2.
\label{Aansatz2}
\end{eqnarray}
It is again straightforward to see that $A_{2}(z)\rightarrow 0$ at the boundary $z=0$, implying that the bulk spacetime asymptotes to AdS at the boundary. Near the boundary $V(z)|_{z\rightarrow 0}=-12/L^2=2 \Lambda$. Further, it can be explicitly checked that $V(z)$ has the same asymptotic form for both $A_{1}(z)$ and $A_{2}(z)$. The parameter $\bar{a}=c/8 \simeq 0.145$ in eq.~(\ref{Aansatz2}) is fixed by demanding the critical temperature of the Hawking/Page (or the dual confinement/deconfinement) phase transition to be around $270 \ \text{MeV}$ at zero chemical potential.\\

Importantly, for $A(z)=- \bar{a} z^2$, the equation for dilaton field in (\ref{metsolution1}) can be solved explicitly
\begin{eqnarray}
\phi(z)=z \sqrt{3 \bar{a}(3+2 \bar{a}z^2)} +3 \sqrt{\frac{3}{2}}\sinh ^{-1} \biggl[ {\sqrt{\frac{2 \bar{a}}{3}}z} \biggr].
\label{phicase2}
\end{eqnarray}
Its near boundary expansion is
\begin{eqnarray}
\phi(z)=6 \sqrt{\bar{a}}z +2/3 \bar{a}^{3/2} z^3+\ldots
\label{phiUVcase2}
\end{eqnarray}
and in terms of $\phi(z)$ the dilaton potential can again be written as
\begin{eqnarray}
V(\phi)=-\frac{12}{L^2}+\frac{\Delta(\Delta-4)}{2}\phi^2(z)+\ldots, \ \ \Delta=3
\label{Vcase2exp}
\end{eqnarray}
We repeat here once more that the phenomenological bottom-up holographic models
we are considering are based on the (implicit) assumption that these phenomenological models are derivable from a consistent truncation of a higher
dimensional string theory. Therefore, the dilaton might be expected to be massless in a 5D truncated gravity theory (like ours) as well. Though, as frequently done, see e.g.~\cite{Colangelo:2007pt,Brunner:2014lya,Brunner:2015oqa,Ballon-Bayona:2017sxa}, the scalar dilaton can be associated to the dual scalar glueball state \cite{Csaki:1998qr} described by $F_{\mu\nu}^2$, which is massive in QCD (or even pure Yang-Mills gauge theory). In fact, whatever the dual of the dilaton would be, QCD(-like) theories are not expected to have any massless physical degrees of freedom in their spectrum, at least away from the chiral limit. As such, it is no surprise that, to our knowledge, phenomenological holographic QCD models are frequently incorporating a massive dilaton as an easy access to a massive scalar glueball, let us refer to \cite{Ballon-Bayona:2017sxa} for a recent discussion and relevant references. Another example with a massive dilaton is \cite{dePaula:2008fp}. Seminal works using bottom-up massive dilatons are \cite{Gubser:2008ny,Gubser:2008yx}, there more in relation to QCD(-like) thermodynamics than to glueball spectra. The mass was related to the (UV) anomalous dimension of $F_{\mu\nu}^2$. One of the considered dilaton potentials in \cite{Gubser:2008ny,Gubser:2008yx} was a simple exponential, as also discussed in \cite{Charmousis:2010zz}, which evidently include a $\phi^2$ term upon expansion. A more refined version of the models considered here in our paper would be to ensure that the quadratic term in the expansion of $V(\phi)$ is related to the anomalous dimension of $F_{\mu\nu}^2$ instead of to the rather simple (and crude) value of $3$ reported now.\\

For completeness, also the $V(\phi)$ of Case 1 gives rise a dilaton mass term in the potential that is consistent with the BF bound for a scalar degree of freedom, see eq.~(\ref{Vcase1exp}). In fact, up to order $\phi^2$, the dilaton potentials of both Cases are identical.

\begin{figure}[t!]
\begin{minipage}[b]{0.5\linewidth}
\centering
\includegraphics[width=2.8in,height=2.3in]{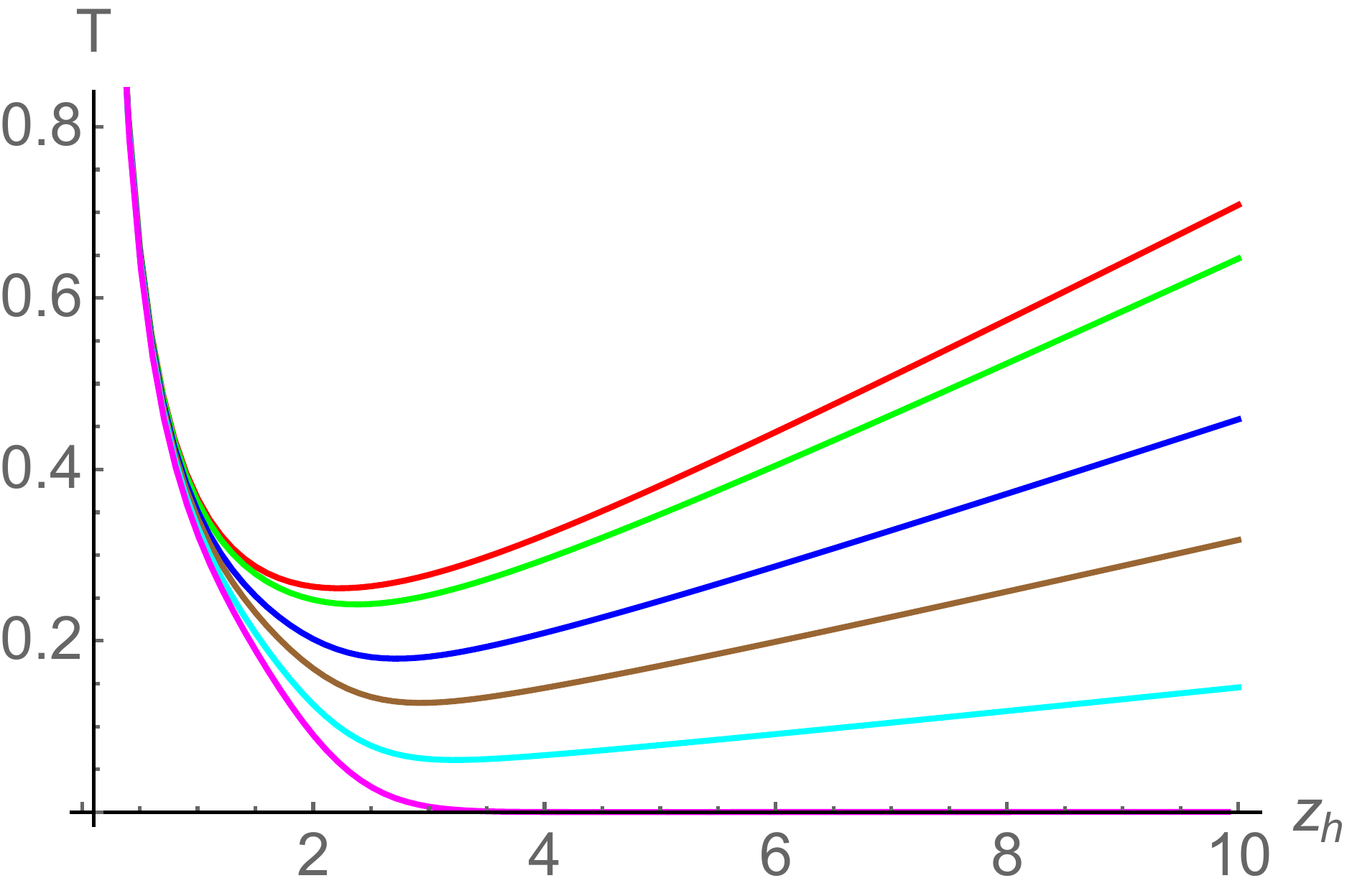}
\caption{ \small $T$ as a function of $z_h$ for various values of the chemical potential $\mu$. Here red, green, blue, brown, cyan and magenta curves correspond to $\mu=0$, $0.2$, $0.4$, $0.5$, $0.6$ and $0.673$ respectively. In units \text{GeV}.}
\label{zhvsTblackholecase2}
\end{minipage}
\hspace{0.4cm}
\begin{minipage}[b]{0.5\linewidth}
\centering
\includegraphics[width=2.8in,height=2.3in]{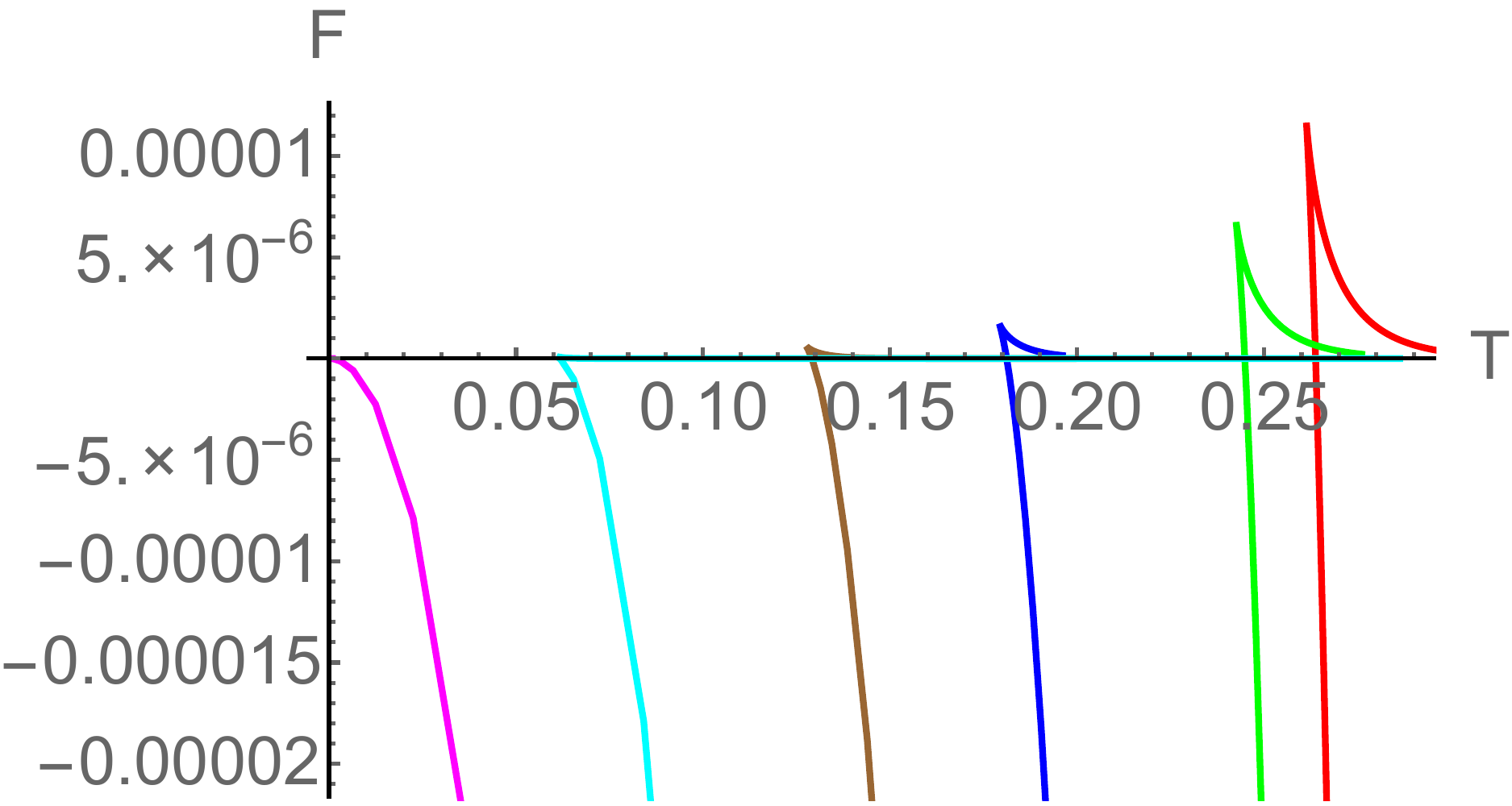}
\caption{\small $F$ as a function of $T$ for various values of the chemical potential $\mu$. Here red, green, blue, brown and cyan curves correspond to $\mu=0$, $0.2$, $0.4$, $0.5$, $0.6$ and $0.673$ respectively. In units \text{GeV}.}
\label{TvsFblackholecase2}
\end{minipage}
\end{figure}
\subsection{Black hole thermodynamics}
The thermodynamic results of the gravity solution with $A(z)$ as in eq.~(\ref{Aansatz2}) are shown in Figures~\ref{zhvsTblackholecase2} and \ref{TvsFblackholecase2}. In this case no small to large black hole phase transition takes place. There are now only two branches in the $(T,z_h)$ plane. The branch with negative slope (small $z_h$) is stable whereas the branch with positive slope (large $z_h$) is unstable. The appearance of this unstable branch however depends on the value of the chemical potential and in particular it ceases to exist for higher chemical potential. This defines a critical chemical potential $\mu_c=0.673 \ \text{GeV}$.  Moreover, the black hole solution does not exist below a certain minimal temperature $T_{min}$. This suggests a phase transition from black hole to thermal-AdS as we decrease the Hawking temperature. The phase transition can be observed from the free energy behaviour, shown in Figure~\ref{TvsFblackholecase2}, where the same normalization as in the previous section is used. We see that the free energy is positive for the unstable branch and becomes negative after some critical temperature $T_{crit}$ along the stable branch, implying a first order Hawking/Page phase transition from thermal-AdS to AdS black hole as the temperature increases. The overall dependence of $T_{crit}$ on $\mu$ is the same as in Figure~\ref{MuvsTcrit}, however with larger $\mu_c=0.673 \ \text{GeV}$~\footnote{A similar kind of Hawking/Page phase transition with a planar horizon has been reported recently in \cite{Zhang:2017tbf}.}.\\

As we will show in the next subsection, this thermal-AdS/black hole phase transition in the gravity side corresponds to the standard confinement/deconfinement phase transition on the dual boundary side. In particular, the thermal-AdS phase corresponds to confinement whereas the black hole phase corresponds to deconfinement. Correspondingly, $T_{crit}$ defines the dual transition temperature. By construction, our result $T_{crit}=0.264 \ \text{GeV}$ at zero chemical potential agrees with its lattice estimate. Moreover, we find that the transition temperature decreases with the chemical potential which is again in qualitative agreement with the lattice results.

\subsection{Free energy of the $q\overline q$ pair}
We now discuss the free energy of the $q\overline  q$ pair  with $A(z)$ as in eq.~(\ref{Aansatz2}). Again, depending upon the background geometry, there can be both connected as well as disconnected string solutions. The equations for the connected (eqs.~(\ref{Fcon}) and (\ref{lengthcon})) and disconnected string (eq.~(\ref{Fdiscon})) remain the same, except $A(z)$ is replaced by eq.~(\ref{Aansatz2}).\\

\begin{figure}[h!]
\begin{minipage}[b]{0.5\linewidth}
\centering
\includegraphics[width=2.8in,height=2.3in]{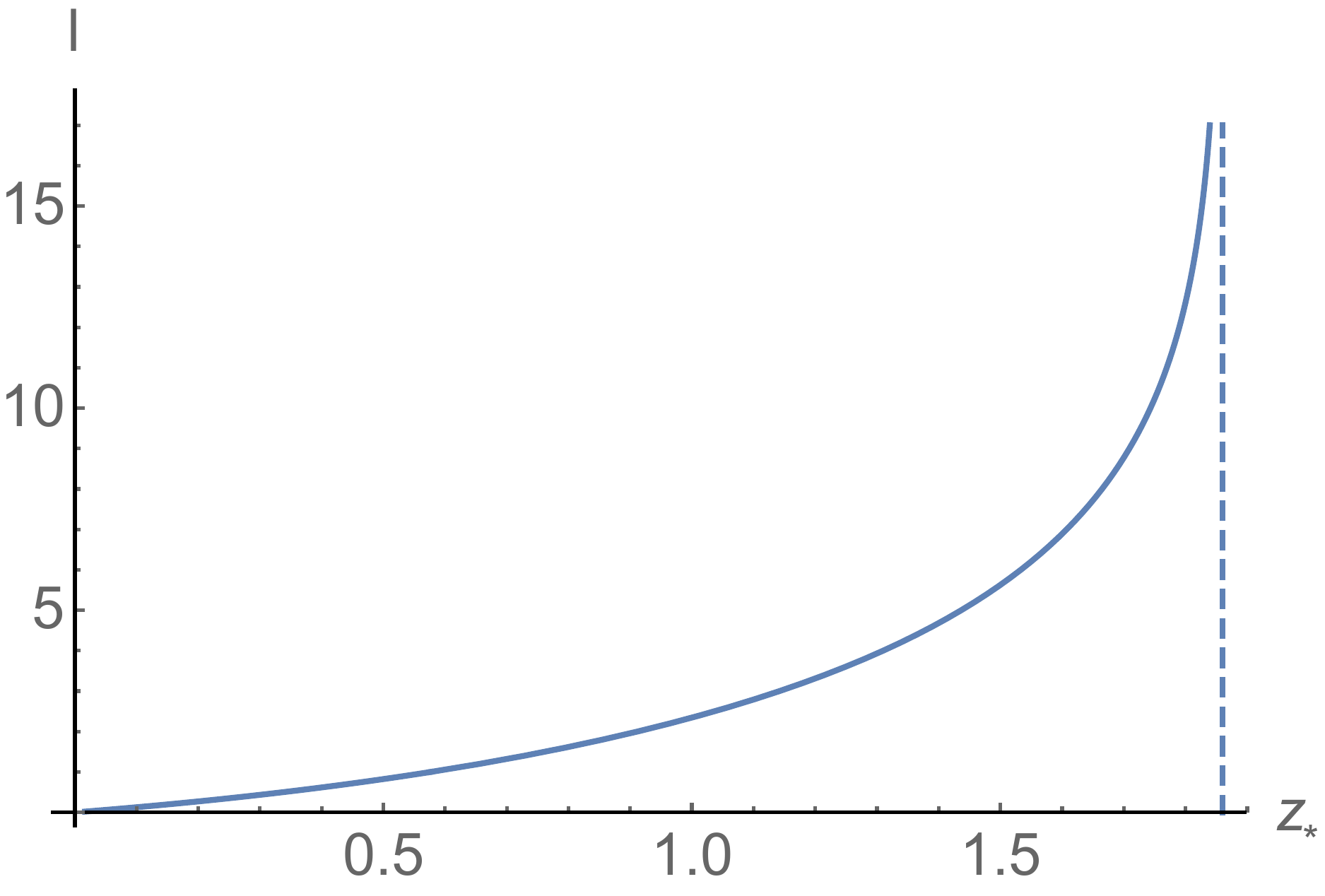}
\caption{ \small $\ell$ as a function of $z_*$ in the thermal-AdS background. In units \text{GeV}.}
\label{lvszsconfdcase2}
\end{minipage}
\hspace{0.4cm}
\begin{minipage}[b]{0.5\linewidth}
\centering
\includegraphics[width=2.8in,height=2.3in]{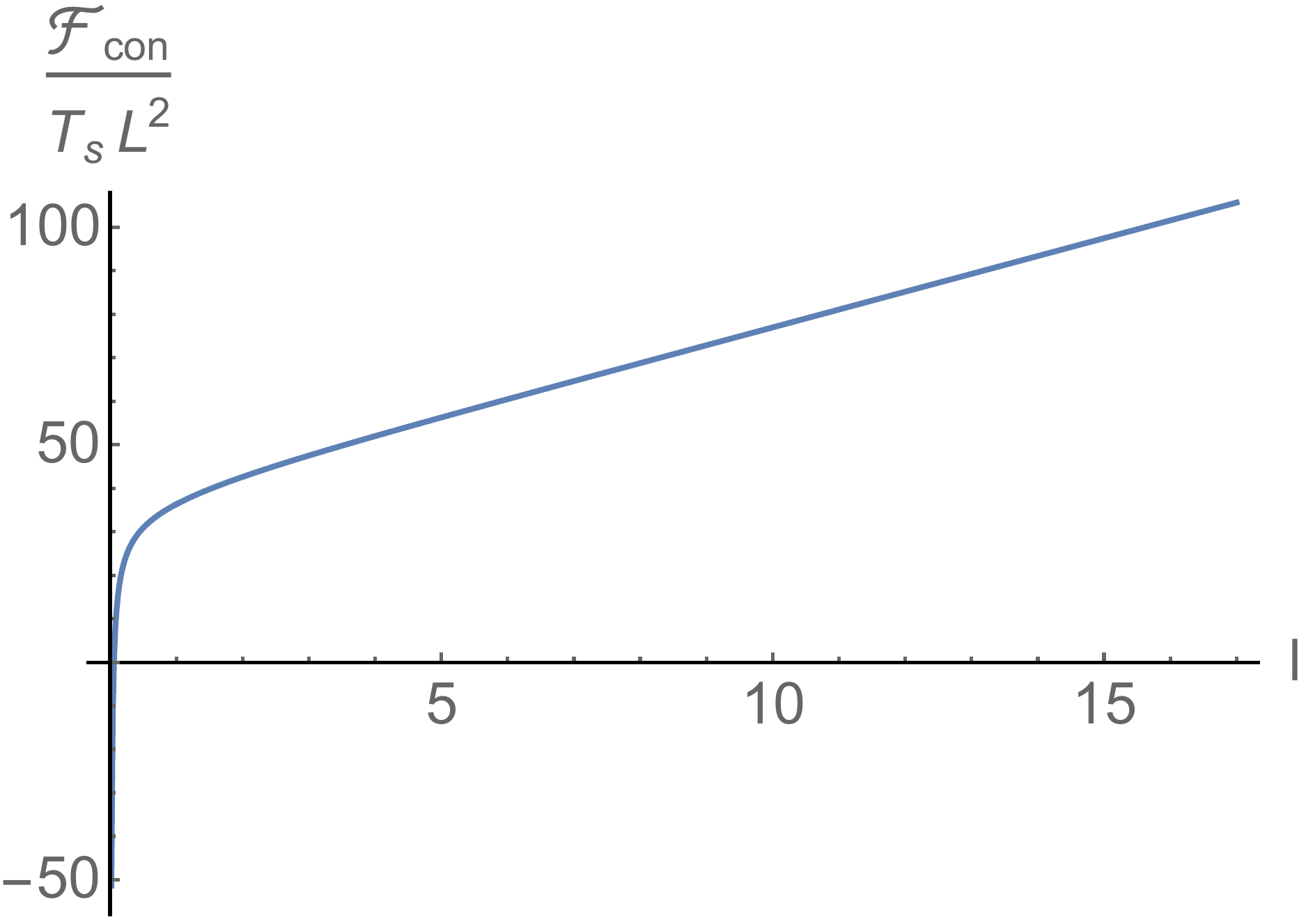}
\caption{\small $\mathcal{F}_{con}$ as a function of $\ell$ in the thermal-AdS background. In units \text{GeV}.}
\label{lvsFconfdcase2}
\end{minipage}
\end{figure}
Let us first examine the free energy in the thermal-AdS background. The results are shown in Figures~\ref{lvszsconfdcase2} and \ref{lvsFconfdcase2}. Again, just as for the small black hole background of the previous section, an ``imaginary wall'' appears and the connected string world sheet does not go deep into the bulk and tries to stay away from the singularity. Importantly, the free energy of the $q\overline q$ pair is again found to be of the Cornell type, $\mathcal{F}_{con}=-\frac{\kappa}{\ell}+\sigma_s \ell$.~\footnote{Again, one can fix the value of open string tension $T_s$ by comparing the lattice QCD result $\sigma_s \approx 1/(2.34)^2 \text{GeV}^2$ with our numerical results. By doing that we find $T_s L^2\simeq 0.1$.}
 This suggests that the $q\overline q$ pair is connected by the string and forms a confined state in the dual boundary theory. However, as opposed to the small black hole background, here $\mathcal{F}_{con}$ is actually always the true minimum of the Nambu-Goto action, which can be easily verified by noticing that the disconnected string solution has an additional IR divergence and therefore is never a relevant solution of Nambu-Goto action. In order to explicitly see this, let us first note the IR (large $z$) expansions of $\phi(z)$ and $A_{s}(z)$,
\begin{eqnarray}
\phi(z)= \sqrt{6}\bar{a} z^2 + \frac{3\sqrt{6}}{2} \ln{z}+\ldots,
\label{phiIRcase2}
\end{eqnarray}
\begin{eqnarray}
& & A_{s}(z)=A(z)+\frac{1}{\sqrt{6}} \phi(z)= -\bar{a} z^2 +\frac{1}{\sqrt{6}} \biggl[ \sqrt{6}\bar{a} z^2 + \frac{3\sqrt{6}}{2} \ln{z}+\ldots\biggr], \nonumber \\
& & \hspace{0.4in} = \frac{3}{2} \ln{z}+\ldots.
\label{phiIRcase2}
\end{eqnarray}
Substituting these expressions into $\mathcal{F}_{discon}$ expression, we have
\begin{eqnarray}
\mathcal{F}_{discon}=\frac{L^2}{\pi \ell_{s}^2}\int^{z_h=\infty} dz \frac{e^{2 A_{s}(z)}}{z^2}=\frac{L^2}{\pi \ell_{s}^2}\int^{z_h=\infty} dz \ \bigl[z+\ldots \bigr]
\end{eqnarray}
where $z_h=\infty$ for thermal-AdS, which makes $\mathcal{F}_{discon}$ divergent. Therefore, $\mathcal{F}_{con}$ is indeed the relevant quantity for the $q\overline q$ free energy and since $\mathcal{F}_{con}$ shows linear confinement for large $q\overline q$ separation length it implies that the boundary theory dual to thermal-AdS corresponds to confinement. Moreover, the Polyakov loop expectation value is  now zero by default. Of course, this discussion is just a straightforward generalization of known facts in the soft wall AdS/QCD models, however, now with a consistent solution of Einstein-Maxwell-dilaton gravity, also displaying the area law for the Wilson loop, a feature missed by the original soft wall model \cite{Karch1012}.\\
\begin{figure}[h!]
\begin{minipage}[b]{0.5\linewidth}
\centering
\includegraphics[width=2.8in,height=2.3in]{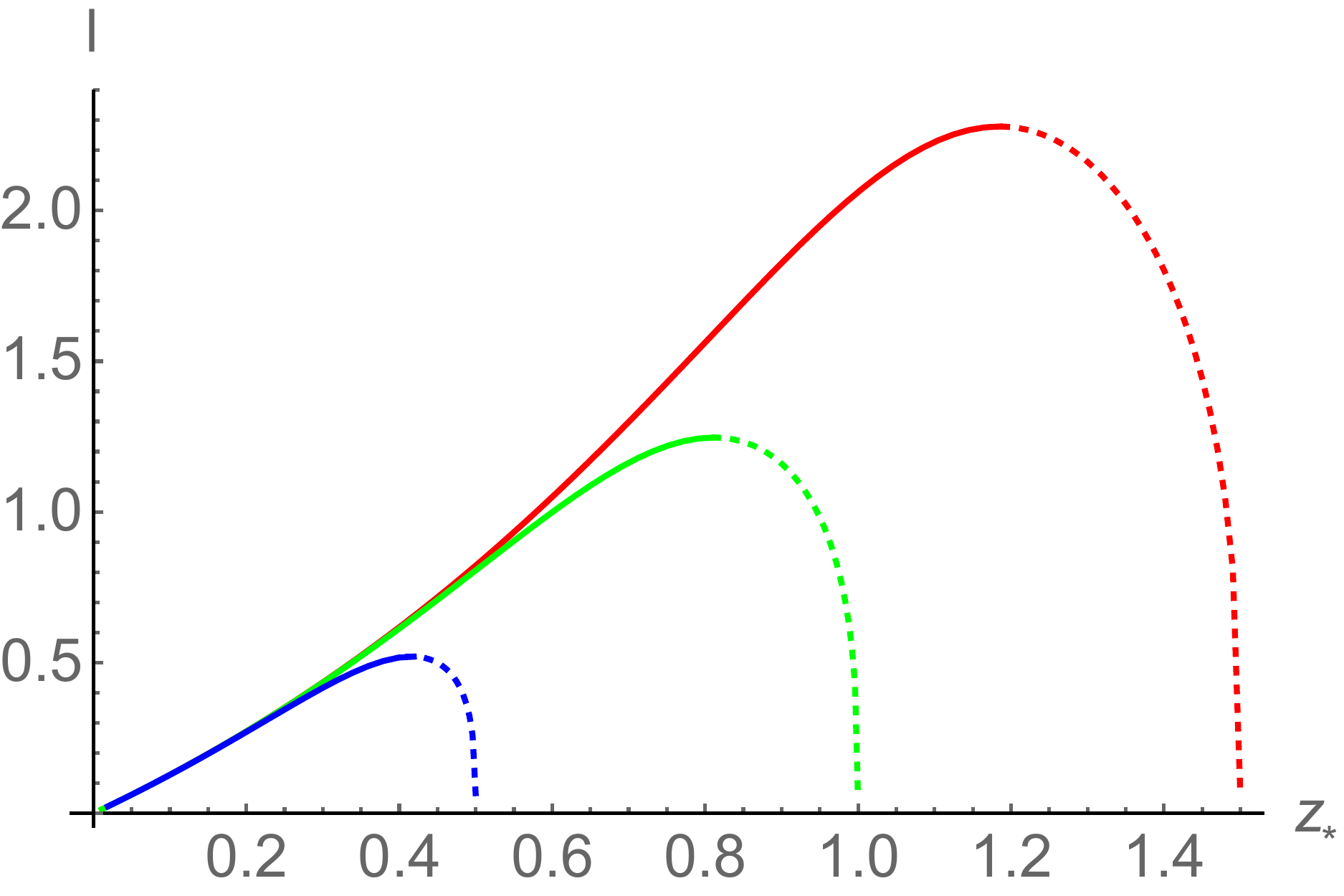}
\caption{ \small $\ell$ as a function of $z_*$ in the AdS black hole background for various values of $z_h$. Here $\mu=0$ and red, green and blue curves correspond to $z_h=1.5$, $1.0$ and $0.5$ respectively. In units \text{GeV}.}
\label{lvszsvszhMu0largecase2}
\end{minipage}
\hspace{0.4cm}
\begin{minipage}[b]{0.5\linewidth}
\centering
\includegraphics[width=2.8in,height=2.3in]{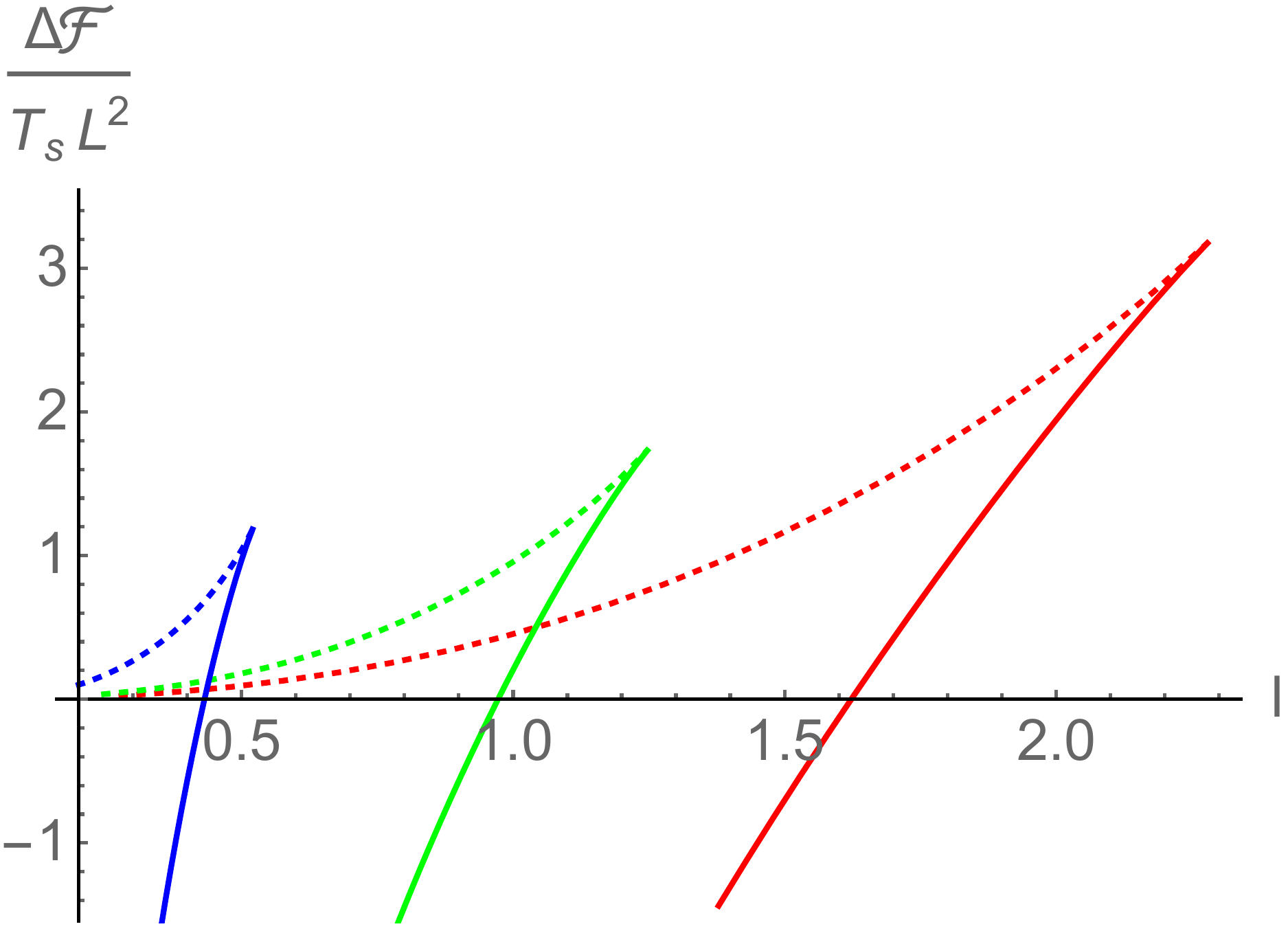}
\caption{\small $\Delta\mathcal{F}=\mathcal{F}_{con}-\mathcal{F}_{discon}$ as a function of $\ell$ in the AdS black hole background for various values of $z_h$. Here $\mu=0$ and red, green and blue curves correspond to $z_h=1.5$, $1.0$ and $0.5$ respectively. In units \text{GeV}.}
\label{lvsFvszhMu0largecase2}
\end{minipage}
\end{figure}

The results for the $q\overline q$ free energy in the AdS black hole background are quite similar to those of the large black hole background of the previous section, and therefore we will be very brief here. The results are summarized in Figures~\ref{lvszsvszhMu0largecase2} and \ref{lvsFvszhMu0largecase2}. For each temperature we again find an $\ell_{max}$ above which the connected string solution does not exist, and a phase transition from connected to disconnected string solution occurs at $\ell_{crit}<\ell_{max}$ as we increase $\ell$. Once again, $\ell_{crit}$ is found to decrease with temperature suggesting the deconfined nature of the $q\overline q$ pair at higher and higher temperatures. With the chemical potential turned on, we find that $\ell_{crit}$ shows no dependence on $\mu$ for temperatures near $T_c$, and only for high temperatures $T\gtrsim 2 T_c$, $\ell_{crit}$ shows a mild dependence on $\mu$.

\subsection{Thermal entropy of the $q\overline q$ pair}
For the thermal AdS background, $\mathcal{F}_{con}$ is the only relevant solution which is independent of temperature. Subsequently, the entropy of the quark pair in the dual confined phase is also zero,
\begin{eqnarray}
S_{confined}=0
\end{eqnarray}
This is in sharp contrast to the \textit{specious-confined} phase of the previous section, which was shown to be dual to a small black hole phase, where the entropy was found to enhance with temperature and to grow to a large magnitude near the transition temperature.\\
\begin{figure}[h!]
\begin{minipage}[b]{0.5\linewidth}
\centering
\includegraphics[width=2.8in,height=2.3in]{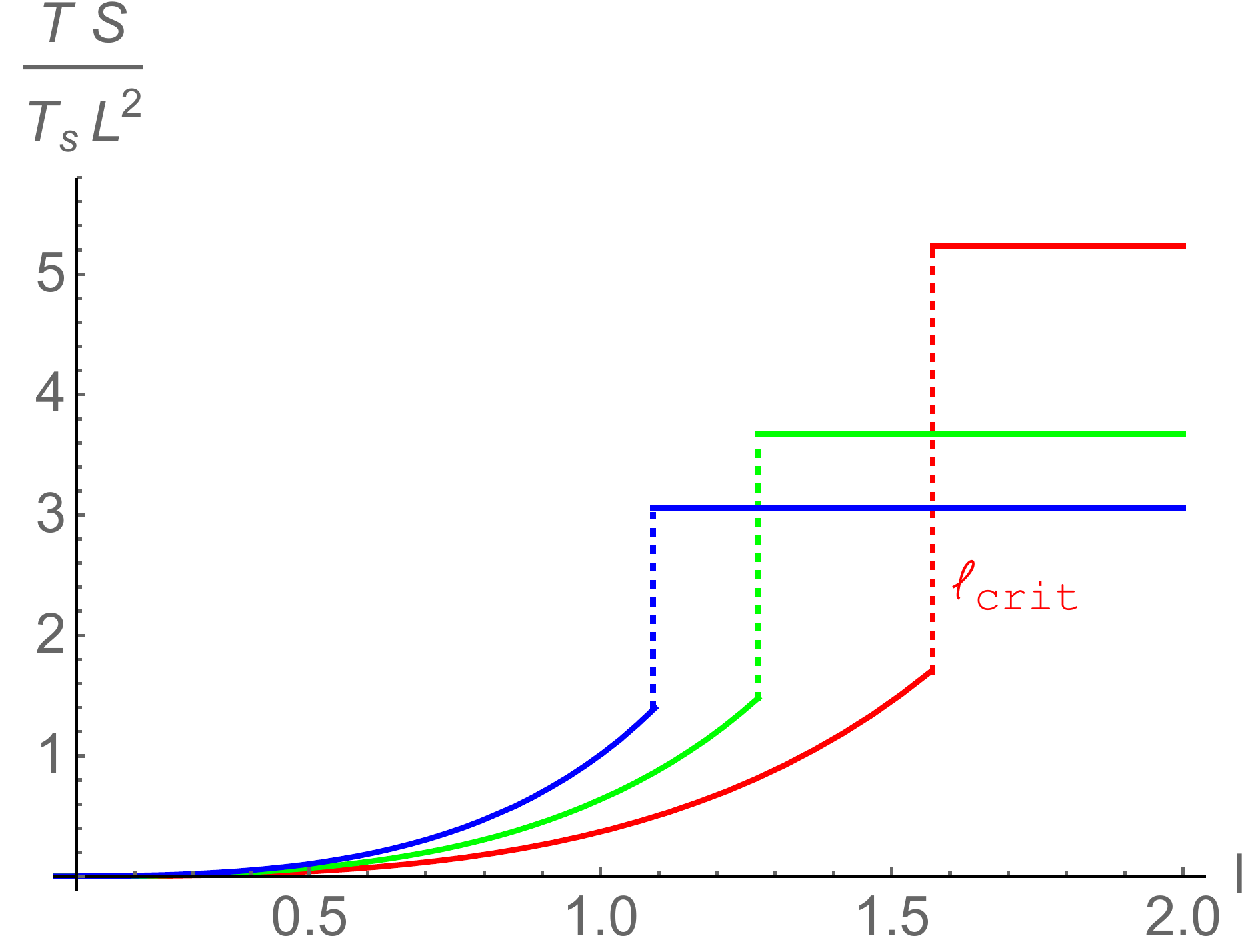}
\caption{ \small Entropy of the $q\overline q$ pair as a function of distance in the deconfined phase for various temperatures. Here $\mu=0$ and red, green and blue curves correspond to $T/T_{crit}=1.1$, $1.2$ and $1.3$ respectively. In units \text{GeV}.}
\label{lvsSvsTMu0case2}
\end{minipage}
\hspace{0.4cm}
\begin{minipage}[b]{0.5\linewidth}
\centering
\includegraphics[width=2.8in,height=2.3in]{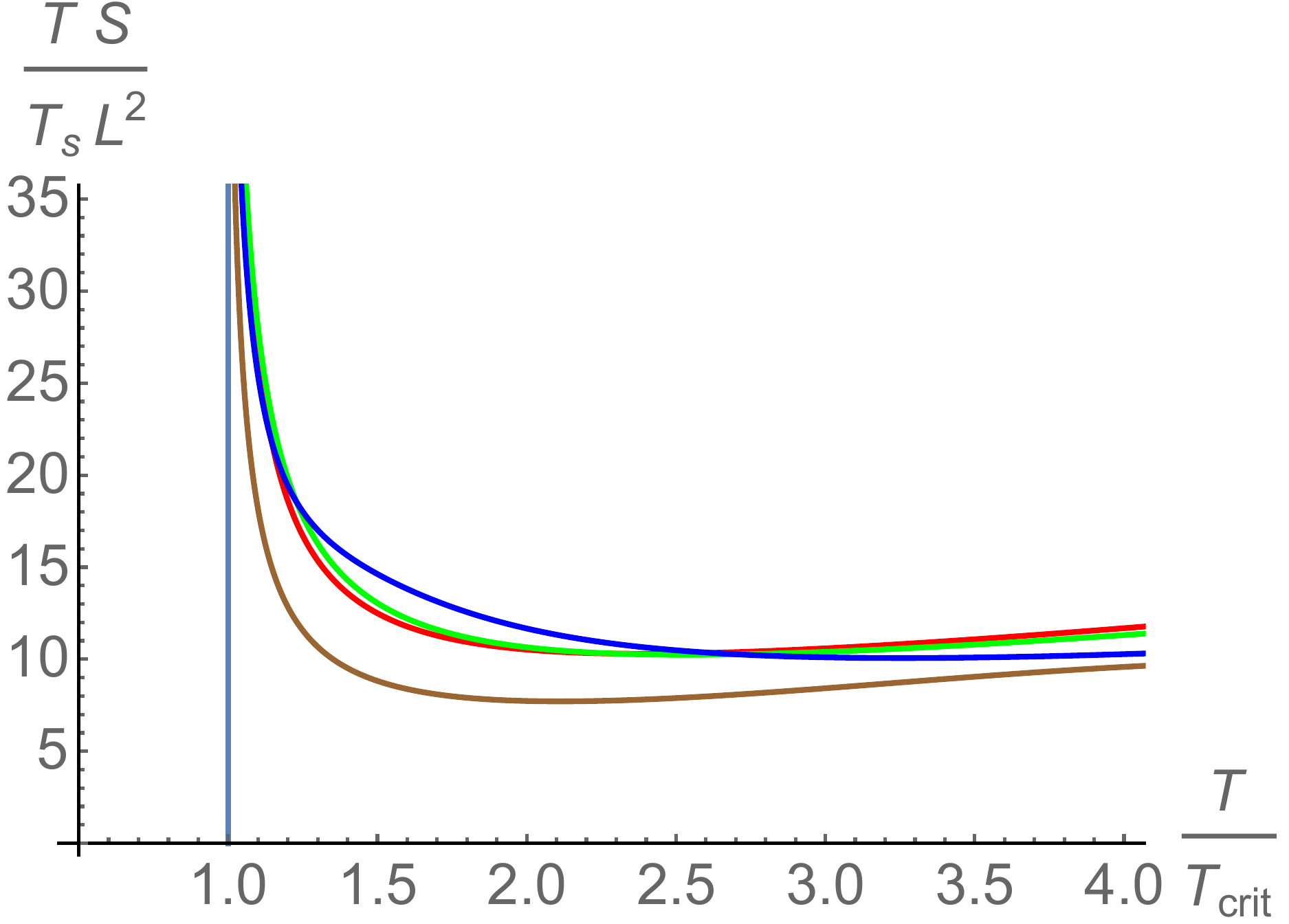}
\caption{\small Entropy of the $q\overline q$ pair as a function of temperature in the deconfined phase for various values of chemical potential $\mu$. Here red, green, blue and brown curves correspond to $\mu=0$, $0.2$, $0.4$ and $0.6$ respectively. In units \text{GeV}.}
\label{TvsSvsMudeconfdcase2}
\end{minipage}
\end{figure}

For the AdS black hole background, we again have two expressions for the entropy. It is given in terms of $\mathcal{F}_{con}$ for small separation length whereas for large separation length it is given in terms of $\mathcal{F}_{discon}$. The main results are shown in Figures~\ref{lvsSvsTMu0case2} and \ref{TvsSvsMudeconfdcase2}~\footnote{The magnitude of $S$ has been suppressed by a factor of 5 in $\ell>\ell_{crit}$ region of Figure~\ref{lvsSvsTMu0case2} in order to make it more readable.}. The essential features of the entropy are again in agreement with lattice QCD. In particular, it can be observed from Figure~\ref{lvsSvsTMu0case2} that the entropy is an increasing function of separation length, which saturates to a constant value at large separations.  We have tested several other forms of the scale factor $A(z)$ as well (apart from $A_{1}(z)$ and $A_{2}(z)$) and found similar results for the deconfined phase. At this point we should emphasize that close results were already described in other holographic models too, both in top-down as well as in bottom-up constructions. For instance, the entropy of a quark pair in $\mathcal{N}=4$ supersymmetric Yang-Mills theory and in bottom-up improved holographic models was shown to increase with inter-quark distance in \cite{Hashimoto:2014fha,Iatrakis:2015sua}. This result therefore seems to be a universal feature of the dual deconfined phase in holographic theories. Moreover, our analysis further suggests that the same behaviour occurs in the presence of chemical potential too. It would be interesting to see whether lattice QCD could predict analogous results in the latter case, at least in the parametric region of small chemical potential which is amenable to some extent to Euclidean lattice simulations.  \\

\section{Speed of sound}
An important thermodynamical observable that is sensitive to the phase transition and attracted a lot of interest of late, both in lattice QCD and holographic QCD, is the speed of sound $C_{s}^2$ \cite{Roessner:2006xn,Boyd:1996bx,Gubser:2008yx,NoronhaHostler:2008ju,Hohler:2009tv,Yang:2017oer}. In conformal theories, the value of speed of sound is fixed $C_{s}^2=1/3$ by conformal invariance. However, in non-conformal field theories it is expected that $C_{s}^2$ depends non-trivially on the temperature. Indeed, lattice QCD has predicted a rapid decrease in the magnitude of $C_{s}^2$ near the deconfinement transition temperature, which approaches its conformal value from below at very high temperatures \cite{Roessner:2006xn,Boyd:1996bx}. Therefore it is of great interest to see how $C_{s}^2$ behaves in our \textit{specious-confined}/deconfined and confined/deconfined phases.
\begin{figure}[h!]
\begin{minipage}[b]{0.5\linewidth}
\centering
\includegraphics[width=2.8in,height=2.3in]{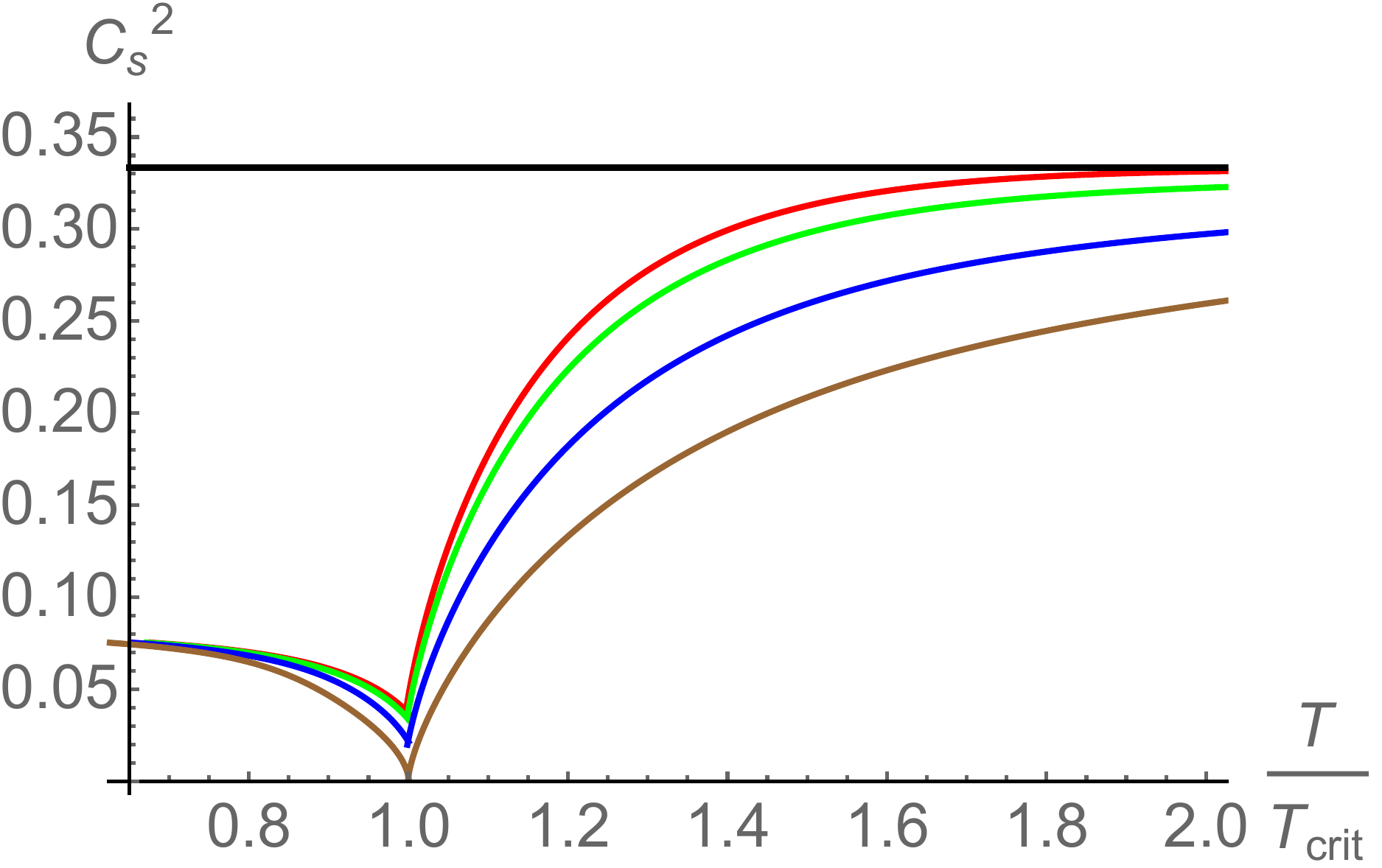}
\caption{ \small $C_{s}^2$ as a function of temperature for various values of $\mu$. Here red, green, blue and brown curves correspond to $\mu=0.0$, $0.1$, $0.2$ and $0.3$ respectively. The solid black line represents the conformal value $C_{s}^2=1/3$. Using $A(z)=A_{1}(z)$. }
\label{soundspeedcase1}
\end{minipage}
\hspace{0.4cm}
\begin{minipage}[b]{0.5\linewidth}
\centering
\includegraphics[width=2.8in,height=2.3in]{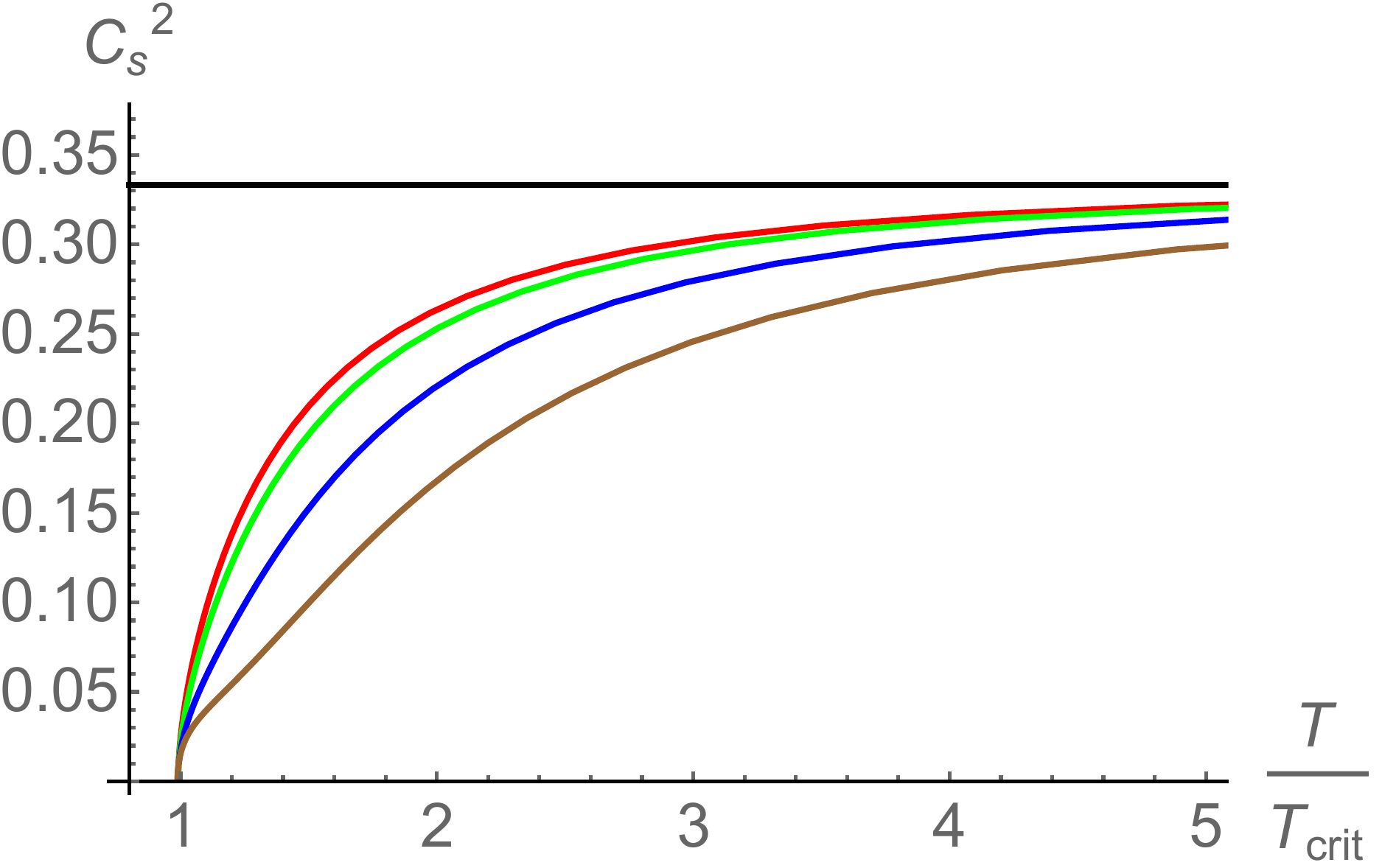}
\caption{\small\small $C_{s}^2$ as a function of temperature for various values of $\mu$. Here red, green, blue and brown curves correspond to $\mu=0.0$, $0.1$, $0.2$ and $0.3$ respectively. The solid black line represents the conformal value $C_{s}^2=1/3$.  Using $A(z)=A_{2}(z)$.}
\label{soundspeedcase2}
\end{minipage}
\end{figure}
\\

The general expression of $C_{s}^2$ in the grand canonical ensemble is given by,
\begin{eqnarray}
C_{s}^2=\frac{S_{BH}}{T \bigl(\frac{\partial S_{BH}}{\partial T}\bigr)_\mu + \mu \bigl(\frac{\partial \rho}{\partial T}\bigr)_\mu }.
\label{soundspeed}
\end{eqnarray}
where $\rho$ is the charge density of the boundary field theory. It can be extracted from the asymptotic expansion of $A_{t}$ and can be expressed in terms of $\mu$ using the gauge/gravity duality mapping. Using eqs.~(\ref{Mueqn}), (\ref{fansatz}) and (\ref{metsolution1}), we get
\begin{eqnarray}
\rho = \frac{\mu}{2 \int_{0}^{z_h} dx \ x e^{c x^2}} = \frac{\mu c}{e^{c z_{h}^2}-1}
\label{density}
\end{eqnarray}
Let us first discuss the results with $A(z)=A_{1}(z)$. Our numerical results for $C_{s}^2$ as a function of temperature for various values of $\mu$ are shown in Figure~\ref{soundspeedcase1}. We find that $C_{s}^2$ sharply decreases near the \textit{specious-confined}/deconfined critical temperature. In particular, $C_{s}^2$ decreases near the critical temperature from the \textit{specious-confined} phase side too. These results are again quite similar to unquenched lattice QCD results, further advocating a close relation between \textit{specious-confined} and genuine confined phases, although that recent lattice results rather predict a smooth minimum of the speed of sound near the transition temperature, instead of the cusp-like variation as found in our model \cite{Borsanyi:2013bia,Bazavov:2014pvz}. Moreover, we find that the effect of $\mu$ is maximal around $T_{crit}$ and shows a mild dependence away from $T_{crit}$. In particular, higher values of $\mu$ try to suppress $C_{s}^2$ near $T_{crit}$. However, at extremely low and high temperatures, $C_{s}^2$ approaches a $\mu$ independent constant value.\\

The results with $A(z)=A_{2}(z)$ are shown in Figure~\ref{soundspeedcase2}. Here too $C_{s}^2$ shows a sharp decrease near $T_{crit}$. The temperature dependence of $C_{s}^2$ in the deconfined phase is quite similar to the previous case, however now, since there is no notion of temperature, we cannot study the temperature dependence of $C_{s}^2$ in the confined phase. Again, higher values of $\mu$ try to suppress $C_{s}^2$ near $T_{crit}$ and at very high temperature $C_{s}^2$ approaches a $\mu$ independent constant value. This seems to be a universal feature of the boundary gauge theory.\\

It is important to investigate the high temperature behaviour of $C_{s}^2$ as it is generally believed that QCD becomes conformal at high temperatures. In our setup, the high temperature limit of $C_{s}^2$ can be obtained by doing a series expansion around $z_h=0$. In this limit, we get
\begin{eqnarray}
C_{s}^2=\frac{1}{3} + \frac{8 z_h A'(0)}{15} +\ldots \nonumber \\
=\frac{1}{3} + \frac{8 A'(0)}{15 \pi T} +\ldots
\label{seriesexp}
\end{eqnarray}
where in the last equation we have used $T=1/(\pi z_h)$.
Since $A'(0)$ is negative for both $A(z)=A_{1}(z)$ and $A(z)=A_{2}(z)$ (and for that matter for any $A(z)$ which leads to an area law of the Wilson loop in the boundary theory \cite{GursoyII}), it suggests that in our models too the speed of sound approaches its conformal value from below at large temperatures. We also have numerically checked for sufficiently large temperatures that $C_{s}^2$ is always less than or equal to $1/3$. This can also be explicitly seen from Figures \ref{soundspeedcase1} and \ref{soundspeedcase2}. Further, for very high temperatures, the effects of chemical potential only appear at the order $T^{-2}$.

\section{Conclusions}
 In this paper, we used the gauge/gravity duality to investigate QCD thermodynamics in relation to its lattice version. We considered an Einstein-Maxwell-dilaton gravity model to study the entropy of a $q\overline q$ pair and the speed of sound in the boundary theory. We first expressed the gravity solution in terms of a scale function $A(z)$ and then considered two different profiles for it, each of which led to various kinds of phase transitions in the gravity side. For $A(z)=A_{1}(z)$, we found a first order small/large black hole phase transition which on the dual boundary side corresponds to a \textit{specious-confined}/deconfined phase transition whereas for $A(z)=A_{2}(z)$ we found a Hawking/Page phase transition which corresponds to the standard confined/deconfined phase transition. In both cases the critical temperature is found to decrease with the chemical potential. We discussed that although this \textit{specious-confined} phase does not strictly correspond to the standard confined phase of quenched QCD, it does exhibit many properties which are quite similar to it and even more alike unquenched QCD, making it an interesting self-consistent holographic model in se. Importantly this \textit{specious-confined} phase, as opposed to the standard confined phase, has a notion of temperature which allows us to study thermal properties of various observables in the confined phase. We studied the free energy and entropy of a $q\overline q$ pair and showed that our holographic model qualitatively describes the known lattice QCD results. In particular, one of the main results of our holographic model is that it predicts a sharp rise in the entropy of a $q\overline q$ pair near the deconfinement transition temperature that can be seen from the \textit{specious-confined} phase side too. We further provided a holographic estimate for the $q\overline q$ entropy when a chemical potential is switched on, which hopefully can be compared to future lattice QCD predictions. For the standard confined phase (dual to thermal-AdS) the entropy of the $q\overline q$ pair is inherently zero. The properties of the deconfined phase are similar for both $A_{1}(z)$ and $A_{2}(z)$ and to other holographic models studied in the literature. We also probed the speed of sound in our constructed  \textit{specious-confined}/deconfined phases and again report results which are qualitatively similar to lattice QCD.\\

We conclude this paper by suggesting some problems which would be interesting to investigate in the future. The most pertinent one would be to investigate the connection between entanglement entropy and the QCD phase diagram. A few works have appeared in this direction especially in the soft wall models \cite{Klebanov:2007ws,Dudal:2016joz,Knaute:2017lll,Ali-Akbari:2017vtb,Wu:2017bza}, however not much has been discussed in self-consistent holographic QCD models like the one presented here. Another interesting problem would be to investigate the effects of a magnetic field on the entropy of a $q\overline q$ pair. As it is by now well known the confinement/deconfinement transition temperature is quite sensitive to the value of magnetic field and one might think that it may thus also affect the $q\overline q$ entropy near the transition temperature. This question is important both from lattice QCD and from holographic point of view. In lattice QCD, it might give non-trivial evidence in favour of the influence of a magnetic field on quarkonium suppression, and in holography, it might provide another scenario where testable and qualitative predictions from the gauge/gravity duality are achievable. However the introduction of a magnetic field will necessarily introduce additional equations on the gravitational side, coming from off-diagonal components of the Einstein equations, for which simple closed analytic results might no longer be obtainable, and one might have to turn to numerically constructed metrics or employ a perturbative expansion in some small parameter(s). We leave these and other issues for future work.

\section*{Acknowledgments}
It is a pleasure to thank L.~Palhares and O.~Andreev for useful discussions. S.~M.~also likes to thank S.~Sugimoto for several helpful discussions and clarifications. We are grateful to O.~Kaczmarek for the permission to take the Figures 1 and 2 from \cite{Kaczmarek:2005zp}. We are grateful for various useful comments made by the anonymous referee. The work of S.~M.~is supported by a PDM grant of KU Leuven.

\end{document}